\newif\ifarxiv
\arxivtrue

\documentclass{article}

\usepackage[round]{natbib}  %
\bibpunct{(}{)}{;}{a}{}{,}  %
\usepackage{hyperref}       %
\usepackage{url}            %
\usepackage{booktabs}       %
\usepackage{amsfonts}       %
\usepackage{amssymb}
\usepackage{nicefrac}       %
\usepackage{microtype}      %
\usepackage{lipsum}
\usepackage{fancyhdr}       %
\usepackage{graphicx}       %

\usepackage{amsmath}
\graphicspath{{media/}}     %

\usepackage{multirow}

\usepackage{bm}
\newcommand{\matr}[1]{\bm{#1}}
\usepackage{multicol}
\usepackage{subcaption}
\usepackage[title]{appendix}%

\ifarxiv
    \usepackage{xcolor}
    \usepackage{PRIMEarxiv}
    \newcommand{\orcid}[1]{}         %
\fi

\begin{document}

\ifarxiv
    \title{A fullwave model of the nonlinear wave equation with multiple relaxations and relaxing perfectly matched layers for high-order numerical finite difference solutions}
    \author{Masashi Sode$^{1,2}$\quad Gianmarco Pinton$^{1,2,\ast}$}
    \maketitle
    \vspace{-0.25in}
    {\centering\small
        $^{1}$Lampe Joint Department of Biomedical Engineering, The University of North Carolina at Chapel Hill, Chapel Hill, NC, United States\par
        $^{2}$Lampe Joint Department of Biomedical Engineering, North Carolina State University, Raleigh, NC, United States\par
        $^{\ast}$Corresponding author: gia@email.unc.edu\par
    }
    \vspace{0.15in}
\else
    \articletype{Paper} %

    \title{A fullwave model of the nonlinear wave equation with multiple relaxations and relaxing perfectly matched layers for high-order numerical finite difference solutions}

    \author{Masashi Sode$^{1, 2}$\orcid{0000-0002-3685-7378}, and Gianmarco Pinton$^{1,2,*}$\orcid{0000-0002-4896-1439}}

    \affil{$^1$Lampe Joint Department of Biomedical Engineering, The University of North Carolina at Chapel Hill, Chapel Hill, NC, United States}

    \affil{$^2$Lampe Joint Department of Biomedical Engineering, North Carolina State University, Raleigh, NC, United States}

    \affil{$^*$Author to whom any correspondence should be addressed.}

    \email{gia@email.unc.edu}
\fi

\ifarxiv\else
\fi

\begin{abstract}
    \textbf{Objective.} Large-scale acoustic simulation underpins the development of ultrasound imaging and therapy, but modeling nonlinearity, frequency-dependent attenuation, and absorbing boundaries in heterogeneous tissue is computationally demanding. We present Fullwave~2, a unified time-domain formulation that represents arbitrary power-law tissue attenuation and perfectly matched layers (PMLs) in a single high-order finite difference framework.\\
    \textbf{Approach.} Attenuation and dispersion are encoded directly into complex coordinate-stretched spatial derivatives through multiple relaxation mechanisms. Because the same mechanism describes both interior attenuation and the absorbing boundary, the convolutional PML (C-PML) becomes a special case of the domain-wide model with no extra computational burden. The formulation preserves the d'Alembertian structure, which allows high-order staggered-grid finite difference stencils optimized for long-distance propagation, and a two-stage C-PML with a transition region ensures numerical stability with multiple relaxations.\\
    \textbf{Main results.} The domain-wide multiple relaxation model reproduces power-law attenuation with less than $5\%$ attenuation error and less than $0.5\%$ phase-velocity error over a 1--20 MHz bandwidth. The two-stage C-PML reaches reflection coefficients below $-49\ \mathrm{dB}$ with a compact $4\lambda$ footprint. Nonlinear propagation is validated against a 1D Burgers solution, with agreement up to the $7^{\text{th}}$ harmonic. The framework is demonstrated on 2D abdominal wall imaging and 3D transcranial rat skull simulations, where it captures complex scattering and aberration artifacts.\\
    \textbf{Significance.} Fullwave~2 unifies nonlinear propagation, arbitrary power-law attenuation, and absorbing boundaries in a single, computationally efficient time-domain formulation, providing an accurate and scalable wave propagation tool for medical ultrasound research.
\end{abstract}

\ifarxiv
\fi

\section{Introduction}

Large-scale acoustic simulations of wave propagation require efficient and accurate algorithms. In recent decades, many finite-difference (FD) methods have been developed for wave propagation in seismology and acoustics \citep{Virieux1984-ch, Virieux1986-gv}. Staggered-grid FD is often associated with higher stability in heterogeneous media with large contrast and is in many cases more accurate than methods implemented on a conventional grid \citep{Zingg2006-hh, Moczo2002-hs, Tan2014-qt}. Given their ease of implementation, speed, parallelizability, and accuracy, finite differences are used extensively in wave propagation in electromagnetism, seismology, and acoustics.

Acoustic wave propagation in ultrasound imaging involves several physical phenomena: diffraction, reflection, scattering, frequency-dependent attenuation, and nonlinearity. A direct simulation of acoustic wave propagation to a target, reflection, and propagation back to a transducer is constrained by two fundamental physical scales: (i) the propagation distance ($\sim 100 \lambda$) and (ii) sub-resolution scatterers ($< \lambda / 10$). To represent both length scales in 3D, simulation fields with a large number of points in space ($\sim 10^9$) are required. Furthermore, numerical methods are challenged by the extremely high dynamic range because the backscattered wave may be 100 dB smaller than the transmitted pulse. For long-range propagation on coarse grids, one of the main challenges for FD is controlling numerical error, usually in the form of space and time dispersion \citep{Tan2014-qt}.

Here we present a formulation of the wave equation that can model nonlinearity and arbitrary attenuation laws, and that can be used with high-order stencils optimized to minimize dispersion and dissipation errors. We refer to this implementation as \emph{Fullwave~2}~\citep{Pinton2021-to}, a domain-wide multiple-relaxation, staggered-grid finite-difference solver designed for ultrasound-scale propagation. Multiple relaxation mechanisms are formulated to model arbitrary power-law attenuation and dispersion in heterogeneous biological tissues within a time-domain ultrasound simulation framework. These relaxation mechanisms, heterogeneous in space, describe tissue attenuation laws and also implement convolutional perfectly matched layers (C-PMLs). The C-PML formulation is extended to support multiple relaxation processes with a two-stage boundary condition, ensuring numerical stability and minimal reflection in heterogeneous lossy media.

The scope of this paper is the formulation, numerical implementation, and validation of the multiple-relaxation framework. The relaxation parameters that reproduce a target power-law attenuation are obtained here by a coarse-to-fine grid search for a set of representative tissue cases. We do not attempt a systematic optimization of these parameters over the full range of attenuation coefficients and exponents found in soft tissue. Fitting a finite set of relaxation mechanisms to a power law is an intrinsically multi-scale problem, because the relaxation parameters span several orders of magnitude in order to cover a broad frequency band, and a finite number of mechanisms can only approximate the fractional frequency dependence of a power law over a bounded band. The resulting optimization landscape is non-convex and poorly conditioned, so an efficient and general calibration method is not straightforward to design. A systematic and automated calibration of the relaxation parameters across the full clinically relevant range of soft-tissue properties is therefore the subject of a dedicated companion paper \citep{Sode2026-rd}, which builds directly on the formulation established here.

We then conduct numerical modeling of wave propagation in 2D complex media in the context of ultrasound imaging applications. Handling the long propagation lengths, heterogeneous media in the human body, large dynamic range, and computational-speed requirements demonstrates the utility of the proposed methods.

\section{Physical model}
\subsection{Multiple relaxation mechanisms as a domain-wide approach for attenuation and perfectly matched layers} \label{sec:multiple_relaxation}

For long propagation distances, the optimization of finite difference stencils is primarily guided by the performance of the discretization of the d'Alembertian operator.
However, there is significant additional complexity in the implementation of attenuation and boundary conditions which does not neatly fit into optimization approaches of the d'Alembertian alone.
Thus, the central idea behind the formulation presented here is to preserve the structure of the d'Alembertian by encoding attenuation and dispersion directly into complex coordinate-stretched spatial derivatives.
The pressure velocity formulation of the d'Alembertian can be described by stretching and adding memory to the differential operators, which we denote as $\nabla_1$ and $\nabla_2$. Then,
\begin{align}
     & \nabla_1 p + \rho \cfrac{\partial \matr{v}}{\partial t} = 0         \label{eq:wave1} \\
     & \nabla_2 \cdot \matr{v} + \kappa \cfrac{\partial p}{\partial t} = 0 \label{eq:wave2}
\end{align}
where $p(\matr{x}, t)$ and $\matr{v}(\matr{x}, t)$ represent the pressure and velocity wavefield at a given position $\matr{x}$ at a given time $t$ respectively,
and $\rho(\matr{x})$ and $\kappa(\matr{x})$ denote the density and the compressibility of the medium at position $\matr{x}$, respectively.
Both $\rho(\matr{x})$ and $\kappa(\matr{x})$, and likewise the relaxation parameters introduced below, are spatially-varying fields. The finite-difference scheme described in Section \ref{sec:staggered_grid} solves the governing equations locally at each grid point using the parameter values assigned to that point, so heterogeneous media are represented by assigning different properties to different grid points, without assuming homogeneity anywhere in the domain.
$\nabla_1$ and $\nabla_2$ in equations (\ref{eq:wave1}) and (\ref{eq:wave2}) are used to denote the complex spatial derivatives
that model attenuation and dispersion while maintaining
the pressure-velocity formulation of the wave equation.

$\nabla_1$ can be written as
\begin{align}
    \cfrac{\partial}{\partial \tilde{x}_1} & = \cfrac{1}{\kappa_{x_1}}\cfrac{\partial}{\partial x} + \sum^N_{\nu=1}\zeta^\nu_{x_1}(t) \ast\cfrac{\partial}{\partial x} \label{eq:relaxation1} \\
    \cfrac{\partial}{\partial \tilde{y}_1} & = \cfrac{1}{\kappa_{y_1}}\cfrac{\partial}{\partial y} + \sum^N_{\nu=1}\zeta^\nu_{y_1}(t) \ast\cfrac{\partial}{\partial y} \label{eq:relaxation2} \\
    \cfrac{\partial}{\partial \tilde{z}_1} & = \cfrac{1}{\kappa_{z_1}}\cfrac{\partial}{\partial z} + \sum^N_{\nu=1}\zeta^\nu_{z_1}(t) \ast\cfrac{\partial}{\partial z} \label{eq:relaxation3}
\end{align}
Here, $\ast$ denotes the temporal convolution operator, and $\zeta^\nu_{x_1}$ is the convolution kernel for the $N$ relaxations, indexed by $\nu$. The convolution kernels are defined as:
\begin{align}
    \zeta^\nu_{x_1}(t) & = - \cfrac{d^\nu_{x1}}{\kappa^2_{x_1}} \: e^{-\left(\frac{d^\nu_{x1}}{\kappa_{x_1}} + \alpha^\nu_{x_1}\right)t} H(t) \label{eq:convolution_kernel_x} \\
    \zeta^\nu_{y_1}(t) & = - \cfrac{d^\nu_{y1}}{\kappa^2_{y_1}} \: e^{-\left(\frac{d^\nu_{y1}}{\kappa_{y_1}} + \alpha^\nu_{y_1}\right)t} H(t)                                 \\
    \zeta^\nu_{z_1}(t) & = - \cfrac{d^\nu_{z1}}{\kappa^2_{z_1}} \: e^{-\left(\frac{d^\nu_{z1}}{\kappa_{z_1}} + \alpha^\nu_{z_1}\right)t} H(t)
\end{align}
Note that $\kappa_{x_1}(\matr{x})$, and $\kappa_{y_1}(\matr{x})$, $\kappa_{z_1}(\matr{x})$ represent a linear scaling of the derivative at position $\matr{x}$.
This scaling parameter modifies the wave velocity in the $x$, $y$ and $z$ directions.
The variables $d^\nu_{x1}$, $d^\nu_{y1}$, and $d^\nu_{z1}$, represent a scaling-dependent damping profile.
$\alpha^\nu_{x_1}$, $\alpha^\nu_{y_1}$, and $\alpha^\nu_{z_1}$ denote a scaling-independent damping profile.
$H(t)$ is the Heaviside or unit step function.
The transformation set for the $\nabla_2$ operator
is identical to that of $\nabla_1$ and the variables associated with this second transformation are denoted by the subscript 2.
These relaxation mechanisms incorporated in $\nabla_1$, $\nabla_2$ are introduced to empirically model attenuation based on observations of the attenuation laws and parameters observed in soft tissue. These mechanisms can be generalized to arbitrary attenuation laws through a process of fitting the relaxation constants.
Numerically, the convolution terms in Eqs. (\ref{eq:relaxation1})-(\ref{eq:relaxation3}) are computed using auxiliary memory variables, as described in Section \ref{sec:convolutional_operator}.

If the coordinate transformation is applied in the same manner to Eqs. (\ref{eq:wave1}) and (\ref{eq:wave2}), i.e. $\nabla_1=\nabla_2$, the system is dispersionless but attenuating, which is the fundamental behavior that governs PMLs. For example, for the case where $N=1$, $\kappa_{x1} = \kappa_{x2} = 1$, and $\alpha_{x1} = \alpha_{x2}=0$, we acquire the classical PML coordinate transformation. For the case where the coordinate transformations are applied in the same manner to Eqs. (\ref{eq:wave1}) and (\ref{eq:wave2}), and $N=1$, $\kappa_{x1}=\kappa_{x2}$, and $\alpha_{x1}=\alpha_{x2}$, we obtain the C-PML coordinate transformation. In the general case when $\nabla_1\neq\nabla_2$, the coordinate stretching introduces dispersion into the system.

This coordinate transformation within the derivatives is similar to what is used in C-PML formulations~\citep{Komatitsch2007-et}.  Originally, C-PMLs were introduced to create absorbing boundary conditions in wave propagation simulation to create an open-ended domain.
They employ a complex coordinate stretching to the wave equation, which effectively attenuates outgoing waves and minimizes the reflection from the boundary. In the C-PML formulation only a single relaxation mechanism ($N=1$) is used as the exact nature of the attenuation doesn't need to be modeled. This attenuation only occurs within the boundary domain. Here the complex spatial derivatives $\nabla_1$ and $\nabla_2$ are written as a scaling of the partial derivative
and a sum of convolutions with $N$ relaxation functions to extend the modeling flexibility for the entire computational domain. Fullwave~2 thus extends C-PML's attenuation and dispersion control capability to a more general multiple relaxation formulation that can accurately model the attenuation physics on the interior of the domain by representing arbitrary attenuation laws. This enables the same general formulation to describe highly flexible attenuation laws that  model heterogeneous parameters to support different behaviors, including power laws, attenuation magnitudes, varying distributions, and power ramps in the boundaries. These behaviors are governed by the free relaxation parameters.

\subsection{Nonlinear wave equation with multiple relaxation mechanisms}
Nonlinear propagation is a fundamental component of acoustics and ultrasound propagation. Here we describe the proposed relaxation framework in the nonlinear regime. Propagation in a nonlinear isotropic, lossless fluid can be described by a system of first order acoustic equations~\citep{huijssen2010iterative}
\begin{equation}
    {\nabla} p + \rho D_{t}\matr{v}=0
    \label{eq:general1}
\end{equation}
\begin{equation}
    {\nabla} \cdot \matr{v}+\kappa D_{t} p=0
    \label{eq:general2}
\end{equation}
where $p$ is the acoustic pressure, $\matr{v}$ is the particle velocity, and $D_t=\partial_t+\matr{v}\cdot{\nabla}$ denotes the total or material derivative. The medium properties are determined by the density $\rho$, and the compressibility $\kappa$. The nonlinear behavior can be a consequence of terms in the material derivative or the equation of state of the medium. The equation of state, up to second-order terms in Eqs.~(\ref{eq:general1}) and (\ref{eq:general2}), can be written as~\citep{huijssen2006constitutive, huijssen2010iterative}
\begin{equation}
    \rho=\rho_0[1+\kappa_0 p]
    \label{eq:state1}
\end{equation}
\begin{equation}
    \kappa=\kappa_0[1+\kappa_0(1-2\beta)p]
    \label{eq:state2}
\end{equation}
\noindent where $\rho_0$ is the equilibrium density, $\kappa_0$ is the equilibrium compressibility, and $\beta=1+B/2A$ is the coefficient of nonlinearity. Here, $B/A$ is the parameter of nonlinearity of the medium, defined from the Taylor expansion of the pressure about the equilibrium state,
\begin{equation}
    p - p_0 = A \frac{\rho'}{\rho_0} + \frac{B}{2} \left( \frac{\rho'}{\rho_0} \right)^2 + \cdots,
    \label{eq:nonlinearity_parameter}
\end{equation}
where $\rho' = \rho - \rho_0$ is the density perturbation. The quantities $A$ and $B$ are the coefficients of this expansion, and their ratio $B/A$ quantifies the strength of the acoustic nonlinearity of the medium. The parameter of nonlinearity is a standard, tabulated property of soft tissue. By neglecting third order products of $p$ and/or $\matr{v}$ and the locally nonlinear terms given by $\matr{v}\cdot\nabla \matr{v}$ and $\matr{v}\cdot\nabla p$, Eqs.~(\ref{eq:general1})-(\ref{eq:state2}) lead to
\begin{equation}
    {\nabla} p + \rho_0[1+\kappa_0 p]\frac{\partial\matr{v}}{\partial t}=0
    \label{eq:general3}
\end{equation}
\begin{equation}
    {\nabla} \cdot \matr{v}+\kappa_0[1+\kappa_0(1-2\beta)p]\frac{\partial p}{\partial t}=0
    \label{eq:general4}
\end{equation}
Note that with additional approximations that remove all locally nonlinear terms this system of equations can be reduced to the well-known Westervelt equation~\citep{aanonsen1984}.
Finally, using the same process that was described in Section \ref{sec:multiple_relaxation}, attenuation can be included in the equation of state by introducing the relaxation mechanisms $ \nabla_1 $ and $ \nabla_2 $, i.e.

\begin{equation}
    {\nabla_1} p + \rho_0[1+\kappa_0 p]\frac{\partial\matr{v}}{\partial t}=0
    \label{eq:general5}
\end{equation}
\begin{equation}
    {\nabla_2} \cdot \matr{v}+\kappa_0[1+\kappa_0(1-2\beta)p]\frac{\partial p}{\partial t}=0
    \label{eq:general6}
\end{equation}
Thus this nonlinear equation maintains the d'Alembertian structure by modifying the temporal derivatives with nonlinear multipliers. Fullwave~2 employs the finite-difference time-domain (FDTD) method to solve the stretched pressure-velocity equations (\ref{eq:general5}) and (\ref{eq:general6}).

\section{Numerical methods}
\subsection{Staggered grid finite difference time domain stencil} \label{sec:staggered_grid}

A variety of finite-difference time-domain (FDTD) operators have been developed to solve the pressure-velocity wave equation. To ensure high numerical stability and accuracy in heterogeneous media with high accuracy,
Fullwave~2 utilizes a previously developed highly optimized staggered-grid finite difference (FD) discretization~\citep{Tan2014-qt}, whose FD operator has 2$M$-th order accuracy in space and fourth-order accuracy in time.
In the 2D case, the staggered-grid finite difference discretization is defined as follows:
\begin{align}
    p_{0,0}^{n}       & = p_{0,0}^{n-1} - \cfrac{1}{\kappa}(D_x^{2M,4} u_{0,0}^{n-1/2} + D_y^{2M,4} v_{0,0}^{n-1/2}) \Delta t \\
    u_{1/2,0}^{n+1/2} & = u_{1/2,0}^{n-1/2} - \cfrac{1}{\rho} D_x^{2M,4} p_{1/2,0}^{n} \Delta t                               \\
    v_{0,1/2}^{n+1/2} & = v_{0, 1/2}^{n-1/2} - \cfrac{1}{\rho} D_y^{2M,4} p_{0, 1/2}^{n} \Delta t
\end{align}
where $\Delta t$ is the time interval, $D_x^{2M,4}$ and $D_y^{2M,4}$ are the spatial derivatives in the $x$ and $y$ directions, respectively, $p_{m,n}^{j}$ is the pressure at the grid point $(m,n)$ at time step $j$, and $u_{i,j}^{n+1/2}$ and $v_{i,j}^{n+1/2}$ are the velocity components at the staggered grid points.
The finite difference operators $D_x^{2M,4}$ and $D_y^{2M,4}$ have 2$M$-th order accuracy in space and fourth-order accuracy in time, where $M$ is the order of the FD operator. The operator in the $x$ direction is defined as:
\begin{align}
    \cfrac{\partial p}{\partial x} & \approx D_x^{2M,4} p_{0,0} = \cfrac{1}{h} \left[ g_{1,1}(p^n_{1/2,1} - p^n_{-1/2,1} + p^n_{1/2,-1} - p^n_{-1/2,-1}) + \sum_{i=1}^{M}g_{i,0}(p^n_{i-1/2,0} - p^n_{-i+1/2,0})  \right]
\end{align}
In this paper, we use $M=4$ for the FD operator, which yields an eighth-order accurate operator in space and fourth-order accurate in time.
Please refer to \citet{Tan2014-qt} for a full description of the derivation and formulation of the staggered-grid finite difference discretization and other possible implementations of these operators.

\subsection{Auxiliary memory variables for convolutional operator}\label{sec:convolutional_operator}
The first term in Eq. (\ref{eq:relaxation1}) can be computed by simply scaling the existing spatial derivative calculation. To compute the second term within the staggered grid finite difference formulation, the convolution operator can be solved numerically using exponential differentiation. This numerical approach simplifies the convolution operation, which theoretically requires storing all the historical time steps, by auxiliary variables. These auxiliary variables act as a memory variable that allows the convolution to be collapsed from a sum for all previous time to an update of the independent variable at the current time based on the memory variable.

The $\nu$ th relaxation in the convolution term at time step $n$ for the derivative coordinate transformation of $\nabla_1$ in the $x$-coordinate is denoted by $(\psi_{x1}^\nu)^n$.
\begin{align}
    (\psi_{x1}^\nu)^n = \left(\zeta_{x1}^\nu \ast \cfrac{\partial}{\partial_x}\right)^n = \int_{0}^{n \Delta t} \left( \cfrac{\partial}{\partial x} \right)^{n \Delta t - \tau} \zeta_{x1}^\nu (\tau) d\tau
\end{align}
Since the grid is staggered, the time integration scheme is defined half a time step between $m\Delta t$ and $(m+1)\Delta t$ so that:
\begin{align}
    (\psi_{x1}^\nu)^n & = \sum_{m=0}^{n-1} \int_{m\Delta t}^{(m+1)\Delta t} \left( \cfrac{\partial}{\partial x} \right) ^{n\Delta t - \tau} \zeta_{x1}^\nu(\tau) d\tau \\
                      & =\sum_{m=0}^{n-1} \left( \cfrac{\partial}{\partial x} \right) ^{n-(m+1/2)} \int_{m \Delta t}^{(m+1)\Delta t} \zeta_{x1}^\nu(\tau) d\tau        \\
                      & =\sum_{m=0}^{n-1} Z_{x1}^\nu (m) \left( \cfrac{\partial}{\partial x} \right) ^ {n-(m+1/2)}
\end{align}
where
\begin{align}
    Z_{x1}^\nu (m)= \int_{m \Delta t}^{(m+1)\Delta t} \zeta_{x1}^\nu(\tau) d\tau
\end{align}
Then, using Eq.~(\ref{eq:convolution_kernel_x}), $Z_{x1}^\nu (m)$ can be computed as:
\begin{align}
    Z_{x1}^\nu (m) & = -\cfrac{d^\nu_{x1}}{\kappa^2_{x1}} \int_{m \Delta t}^{(m+1)\Delta t} e^{-\left(\frac{d^\nu_{x1}}{\kappa_{x1}} + \alpha^\nu_{x_1}\right)\tau} d\tau \\
                   & = a^\nu_{x1} e^{-\left(\frac{d^\nu_{x1}}{\kappa_{x1}} + \alpha^\nu_{x_1}\right)m\Delta t}
\end{align}
where
\begin{align}
    a^\nu_{x1} = \cfrac{d_{x1}^{\nu}}{\kappa_{x1}(d_{x1}^{\nu} + \kappa_{x1} \alpha_{x1}^{\nu})} (b_{x1}^{\nu} - 1) \label{eq:relaxation_a}
\end{align}
and
\begin{align}
    b_{x1}^{\nu} = e^{-(d_{x1}^{\nu} / \kappa_{x1} + \alpha_{x1}^{\nu})\Delta t} \label{eq:relaxation_b}
\end{align}

Computationally, the convolution is efficient because the memory variable $(\psi_{x1}^{\nu})^n$ requires only a recursive update (Eq.~(\ref{eq:relaxation_psi})) rather than storage of the full time history.
\begin{align}
    (\psi_{x1}^{\nu})^n = b_{x1}^{\nu} (\psi_{x1}^{\nu})^{n-1} + a_{x1}^{\nu} \left(\cfrac{\partial}{\partial x} \right)^{n-1/2} \label{eq:relaxation_psi}
\end{align}
Then, each spatial derivative for the $x$-coordinate in the $\nabla_1$ in Eq. (\ref{eq:relaxation1}) can be replaced by
\begin{align}
    \cfrac{\partial}{\partial \tilde{x}} = \cfrac{1}{\kappa_{x1}} \cfrac{\partial}{\partial x} + \sum_{\nu=1}^{N} \psi_{x1} ^ {\nu} \label{eq:relaxation4}
\end{align}
Identical calculations will lead to equivalent expressions for $\frac{\partial}{\partial \tilde{y}}$ and $\frac{\partial}{\partial \tilde{z}}$ in the $\nabla_1$ operator in Eq.~(\ref{eq:relaxation1}).
The isotropic relaxation model contains $2 + 4N$ parameters to model attenuation and dispersion, where $N$ is the number of relaxation mechanisms.
The parameters are $\kappa_{1}$, $\kappa_{2}$, $d^\nu_{1}$, $d^\nu_{2}$, $\alpha^\nu_{1}$, and $\alpha^\nu_{2}$ for the first and second relaxation mechanisms, where $\nu$ is the relaxation index, and subscript 1 and 2 correspond to the complex spatial derivatives $\nabla_1$ and $\nabla_2$.
The dispersion and attenuation in this system can be quantified with the dispersion relationship describing the wavenumber, $k$.
\begin{align}
    k = \cfrac{\omega}{c} \left( \cfrac{1}{\kappa_{x1} \kappa_{x2}} - \cfrac{\gamma_1}{\kappa_{x2}} - \cfrac{\gamma_2}{\kappa_{x1}} + \gamma_1 \gamma_2 \right) ^ {-\frac{1}{2}} \label{eq:dispersion_relations}
\end{align}
where $c=1/\sqrt{\kappa \rho}$ and
\begin{align}
    \gamma_1 = \sum_{\nu=1}^{N} \cfrac{d^\nu_{x1}}{\kappa^2_{x1}} \left( \cfrac{1}{d_{x1}^\nu / \kappa_{x1} + \alpha_{x1}^{\nu} + i \omega}  \right) \label{eq:gamma1}
    \\
    \gamma_2 = \sum_{\nu=1}^{N} \cfrac{d^\nu_{x2}}{\kappa^2_{x2}} \left( \cfrac{1}{d_{x2}^\nu / \kappa_{x2} + \alpha_{x2}^{\nu} + i \omega}  \right)
\end{align}
If $\gamma_1=\gamma_2=0$, the system is dispersionless and reduces to the lossless wave equation with $k=\omega/c$.
We will use this wavenumber to define the analytical attenuation and dispersion laws in the optimization process.
The derivation of this dispersion relation is provided in the Supplementary Information \ref{appendix:dispersion_relation}.

\subsection{C-PML boundary and transition layer settings}

\begin{figure}[t]
    \centering
    \includegraphics[width=1.0\linewidth]{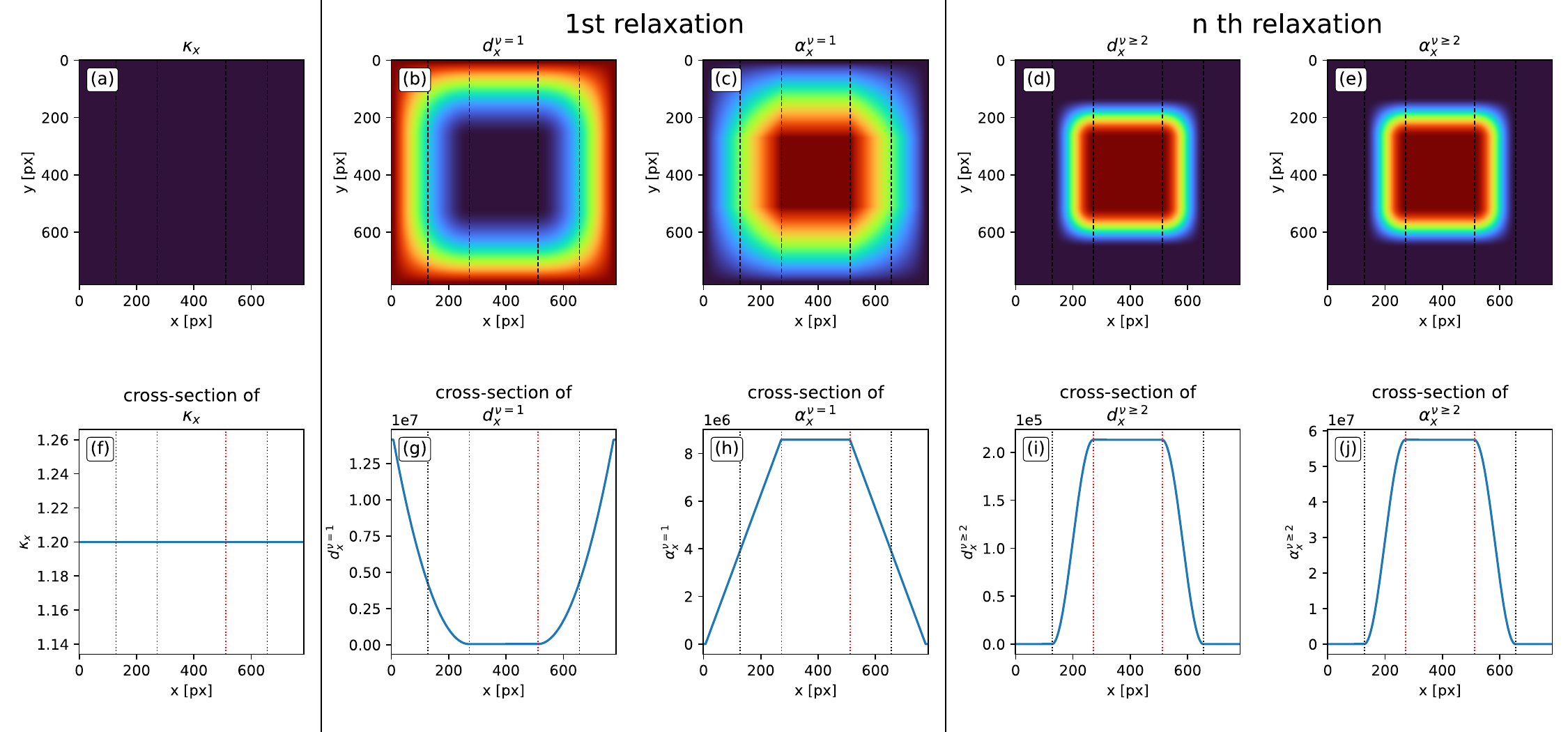}
    \caption{Two stage C-PML and transition layer parameter maps and their cross-sections.
        First row: relaxation parameter maps in 2D simulation domain. It shows the relaxation parameter maps for (a) scaling parameter $\kappa$, (b) damping coefficient $d^{\nu=1}$, and (c) scaling-independent damping coefficient $\alpha^{\nu=1}$ in the C-PML region. (d, e) Transition layer map of damping coefficient $d^{\nu\geq2}$ and scaling-independent damping coefficient $\alpha^{\nu\geq2}$, respectively.
        Second row: cross-section of the relaxation parameters in the x direction. The cross-sections describe the transition from the multiple relaxation model to the single relaxation C-PML.
        (f - h) show the cross-sections for $\kappa$, $d^{\nu=1}$, and $\alpha^{\nu=1}$ in the C-PML region, respectively.
        (i, j) show the cosine transition function for $\nu \geq 2$ to avoid the discontinuity in the transition layer.
    }
    \label{fig:pml_figures}
\end{figure}

The proposed formulation can represent both the multiple relaxation model for modeling arbitrary attenuation laws and the C-PML absorbing boundary condition within a unified framework.
We form an absorbing boundary condition for the first relaxation mechanisms ($\nu=1$).
As described in Section \ref{sec:multiple_relaxation}, C-PML is a special case of the multiple relaxation model with a single relaxation mechanism ($N=1$).
In addition to the C-PML region, we introduced a transition layer that gradually converts the multiple relaxation model into a single relaxation model in the transition region.
The C-PML was originally developed for the case of single relaxation mechanism with $N = 1$ in Eq. (\ref{eq:relaxation4}), meaning it cannot be applied directly to the multiple relaxation model with $N \geq 2$.
This induces numerical instabilities if the relaxation parameters for the relaxation index $\nu \geq 2$ are kept constant within the C-PML region.
To address this issue, we introduced a transition layer that smoothly adjusts the relaxation parameters for the relaxation index $\nu \geq 2$, which gradually reduces their influence to zero at the inner edge of the C-PML region.
Table \ref{tab:pml_parameters} summarizes the C-PML and transition layer parameters used in the boundary region and Figure \ref{fig:pml_figures} illustrates the C-PML and transition layer parameter maps and their cross-sections.

As described in Table \ref{tab:pml_parameters}, we used three types of transition functions to define the C-PML and transition layer parameter distributions. These functions adjust the initial values of the relaxation parameters from the inner edge to the outer edge of the C-PML and transition layers.
In the C-PML region, the scaling-dependent damping coefficient $d^{\nu=1}$ increases polynomially from the inner edge to the outer edge of the C-PML region to effectively attenuate outgoing waves (Fig.~\ref{fig:pml_figures}b and g). Similarly, the scaling-independent damping coefficient $\alpha^{\nu=1}$ decreases linearly from the inner edge to the outer edge of the C-PML region (Fig.~\ref{fig:pml_figures}c and h). The scaling parameter $\kappa$ is kept constant throughout the C-PML region to minimize impedance discontinuities at the PML interface (Fig.~\ref{fig:pml_figures}a and f).
These C-PML settings are adapted from those used in Komatitsch's work~\citep{Komatitsch2007-et}, and the target values for $\tilde{d}$ and $\tilde{\alpha}$ at the outer edge of the C-PML layer are defined accordingly (see Table \ref{tab:pml_parameters}).

The polynomial function for the scaling-dependent damping coefficient $d^{\nu=1}(x)$ is defined as:
\begin{align}
    g_{\text{polynomial}}(x) & = (x/L_{\text{B}})^N, \quad x \in [0, L_{B}] \label{eq:polynomial_transition}
\end{align}
The linear transition function for the scaling-independent damping coefficient $\alpha^{\nu=1}$ is defined as:
\begin{align}
    g_{\text{linear}}(x) & = x/L_{\text{B}}, \quad x \in [0, L_{B}] \label{eq:linear_transition}
\end{align}
As for the transition layer, the scaling-dependent damping coefficient $d^{\nu\geq2}$ and the scaling-independent damping coefficient $\alpha^{\nu\geq2}$ decrease using a cosine transition function from their initial values within the simulation domain to zero at the inner edge of the C-PML region (Fig.~\ref{fig:pml_figures}d, e, i, and j). The scaling parameter $\kappa$ is also kept constant in the transition layer (Fig.~\ref{fig:pml_figures}a and f).
The target values for $d^{\nu\geq2}$ and $\alpha^{\nu\geq2}$ at the outer edge of the transition layer are set to $0.0$, effectively eliminating the additional relaxation mechanisms in the C-PML region.
The cosine transition function for the relaxation parameters for $\nu \geq 2$ is defined as:
\begin{align}
    g_{\text{cosine}}(x) & = \cfrac{1}{2} \left(1 - \cos\left(\pi \cfrac{x}{L_{B}}\right)\right), \quad x \in [0, L_{B}] \label{eq:cosine_transition}
\end{align}
The C-PML and transition layer parameter distributions in the boundary region are defined by inserting each of these transition functions (Eqs.~(\ref{eq:polynomial_transition}), (\ref{eq:linear_transition}), and (\ref{eq:cosine_transition})) into the following general formula:
\begin{align}
    X(x) = (\tilde{X} - X_0) g(x) + X_0, \quad x \in [0, L_{B}]
\end{align}
where $X$ represents the relaxation parameters $d$ and $\alpha$, $X_0$ is the initial value at the inner edge of the C-PML layer, $\tilde{X}$ is the target value at the outer edge of the assigned layer, $L_{\text{B}}$ is the thickness of the assigned layer $B \in \{\text{C-PML}, \text{Transition}\}$, and $g(x)$ is the transition function used for each parameter as described above.
The dispersion relation in the C-PML was kept constant relative to the interior of the domain to minimize any impedance discontinuities at the PML interface by keeping $\kappa$ constant.

C-PML layer and the transition layer are combined to form the boundary region.
For the thickness $L_{\text{C-PML}}$ in the C-PML, we used $L_{\text{C-PML}} = L_{\text{Transition}} + L_{\text{Bound}}$, where $L_{\text{Transition}}$ is the thickness of the transition layer and $L_{\text{Bound}}$ is the thickness of the additional boundary layer for the C-PML.
$L_{\text{Transition}}$ and $L_{\text{Bound}}$ are user-defined parameters and can be adjusted based on the specific simulation requirements.
We investigated the effect of $L_{\text{Transition}}$ and $L_{\text{Bound}}$ on the reflection coefficient in Section \ref{sec:ref_coeff_vs_pml_thickness}.
\begin{table}[t]
    \centering
    \caption{C-PML parameters used in the boundary region.}
    \begin{tabular}{c|c|c|c}
        \hline
        Parameter           & Initial value                     & Target value                                                & Transition function                      \\ \hline
        $\kappa$            & $\kappa_0 = \kappa(x)$            & $\tilde{\kappa} = \kappa_0$                                 & -                                        \\
        $d^{\nu=1}$         & $d_0 = d^{\nu=1}(x)$              & $\tilde{d} = - (N + 1) c_p \log(R_c) / (2L_{\text{C-PML}})$ & Polynomial $g_{\text{polynomial}}(x)$    \\
        $\alpha^{\nu=1}$    & $\alpha_0 = \alpha^{\nu=1}(x)$    & $\tilde{\alpha} = 0.0$                                      & Linear transition $g_{\text{linear}}(x)$ \\
        $d^{\nu\geq2}$      & $d_0 = d^{\nu\geq2}(x)$           & $\tilde{d} = 0.0$                                           & Cosine transition $g_{\text{cosine}}(x)$ \\
        $\alpha^{\nu\geq2}$ & $\alpha_0 = \alpha^{\nu\geq2}(x)$ & $\tilde{\alpha} = 0.0$                                      & Cosine transition $g_{\text{cosine}}(x)$ \\
        \hline
    \end{tabular}
    \label{tab:pml_parameters}
\end{table}

\subsection{Simulation implementation details}
The Fullwave~2 simulation is implemented in CUDA with a Python wrapper.
This enables efficient high performance computation on GPU while providing a user-friendly interface for setting up and running simulations.
The experiments described in this paper are performed on a workstation with an Intel i9-13900K, 128 GB of RAM, and a single NVIDIA RTX 4090 GPU with 24 GB of memory. The operating system is Ubuntu 22.04 LTS with CUDA 12.6 and Python 3.12.
We used a Courant-Friedrichs-Lewy (CFL) condition of 0.4 and a points-per-wavelength (PPW) value of 12 for all simulations in this paper to ensure numerical stability and minimize numerical dispersion, respectively.

\section{Validation \& Benchmarks}

\subsection{Attenuation and dispersion accuracy vs analytical model}

To model arbitrary power-law attenuation and dispersion in biological tissues, the relaxation parameters in the multiple relaxation model need to be optimized to fit the desired attenuation and dispersion characteristics.
The optimization process involves minimizing the difference between the analytical attenuation and dispersion laws and those predicted by the multiple relaxation model.
The optimization process involves minimizing the following cost function:
\begin{align}
    J = w_\alpha \cdot \text{NMSE}\left(\alpha_{\text{model}}(f), \alpha_{\text{target}}(f)\right) + (1-w_\alpha) \cdot \text{NMSE}\left(c_{\text{model}}(f), c_{\text{target}}(f)\right)
    \label{eq:relaxation_parameter_optimization}
\end{align}
where $\alpha_{\text{model}}(f)$ and $c_{\text{model}}(f)$ are the attenuation and phase velocity predicted by the multiple relaxation model at frequency $f$ derived from the dispersion relation in Eq. (\ref{eq:dispersion_relations}), while $\alpha_{\text{target}}(f)$ and $c_{\text{target}}(f)$ are the target attenuation and phase velocity based on the desired power-law characteristics. The weight $w_\alpha$ can be adjusted to prioritize the fitting of attenuation or dispersion. $w_\alpha$ is set to 0.1 in this study to balance the fitting of both attenuation and dispersion.
NMSE denotes the normalized mean square error, given by the ratio of the mean squared error to the variance of the target:
\begin{align}
    \text{NMSE}(x, y) = \dfrac{mean\left(\left( x - y \right)^2\right)}{Var\left( y \right)}
\end{align}
Here the mean and the variance are defined as in \citep{Ross2014}, while the NMSE is their ratio. Normalizing the mean squared error by the variance of the target makes the metric dimensionless and independent of the absolute scale of the fitted quantity. This allows the attenuation error and the phase-velocity error, which have different physical units and magnitudes, to be combined into the single weighted cost function of Eq.~(\ref{eq:relaxation_parameter_optimization}) on a common footing, so that the weight $w_\alpha$ represents a meaningful trade-off. An NMSE of zero corresponds to a perfect fit.
In this paper, we used power law attenuation of the form $\alpha(f) = \alpha_0 f^y$ as the target attenuation law, where $\alpha_0$ is the attenuation coefficient in dB/cm/(MHz$^{y}$), and $y$ is the power-law exponent, and the corresponding phase velocity is derived from the Kramers-Kr\"{o}nig relation.
We have performed a coarse-to-fine grid search constrained by bounds to optimize the relaxation parameters.
Coarse-to-fine grid search was performed by first searching a wide range of relaxation parameters with a coarse grid, followed by multiple finer searches around the best parameters found in the previous search.
The bounds were set as listed in Table \ref{tab:relaxation_parameters}, and were reduced by 30\% of the previous search range in each dimension for each subsequent search.

The parameters except for $\kappa_{x1}$ and $\kappa_{x2}$ were searched in the logarithmic scale to cover a wide range of values efficiently.
The target attenuation parameters were set to $\alpha_0 = 0.5 \ \text{dB/cm/MHz}$ and $y = 1.0$ with two relaxation mechanisms. A power-law exponent of $y=1$ is representative of soft tissue, for which the attenuation exponent is generally close to unity \citep{Wells1975-lq, Parker2022-yz}. Two relaxation mechanisms were chosen as they provide a good balance between fitting accuracy and computational cost for ultrasound applications.
The obtained relaxation parameters are listed in Table \ref{tab:relaxation_parameters}.
\begin{table}[t]
    \centering
    \caption{Optimized relaxation parameters for two relaxation mechanisms targeting $\alpha_0 = 0.5 \ \text{dB/cm/MHz}$ and $y = 1.0$ and their bounds.}
    \begin{tabular}{cccc}
        \toprule
        Relaxation parameters & Optimal value        & Lower bound & Upper bound \\
        \midrule
        $\kappa_{x1}$         & 0.835                & 0.8         & 1.2         \\
        $\kappa_{x2}$         & 1.19                 & 0.8         & 1.2         \\
        $d_{x1}^{\nu=1}$      & $9.82 \times 10^{5}$ & $10^{2}$    & $10^{10}$   \\
        $d_{x2}^{\nu=1}$      & $4.40 \times 10^{4}$ & $10^{2}$    & $10^{10}$   \\
        $\alpha_{x1}^{\nu=1}$ & $2.83 \times 10^{8}$ & $10^{5}$    & $10^{12}$   \\
        $\alpha_{x2}^{\nu=1}$ & $8.35 \times 10^{6}$ & $10^{5}$    & $10^{12}$   \\
        $d_{x1}^{\nu=2}$      & $3.26 \times 10^{6}$ & $10^{2}$    & $10^{10}$   \\
        $d_{x2}^{\nu=2}$      & $1.68 \times 10^{5}$ & $10^{2}$    & $10^{10}$   \\
        $\alpha_{x1}^{\nu=2}$ & $3.54 \times 10^{9}$ & $10^{5}$    & $10^{12}$   \\
        $\alpha_{x2}^{\nu=2}$ & $5.10 \times 10^{7}$ & $10^{5}$    & $10^{12}$   \\
        \bottomrule
    \end{tabular}
    \label{tab:relaxation_parameters}
\end{table}

For the other attenuation parameters, we have modulated the relaxation parameters based on the optimized parameters for $\alpha_0=0.5$ and $y=1.0$.
This modulation is performed by scaling the relaxation parameters $d_x^{\nu=1}$ to achieve the desired attenuation characteristics without re-optimizing the parameters from scratch. The modulation is performed as follows:
\begin{align}
    d_{x}^{\nu=1}(\alpha_0) = d_{x}^{\nu=1} ({\alpha_{\text{ref}}}) \cdot \left( \cfrac{\alpha_0}{\alpha_{\text{ref}}} \right) \label{eq:relaxation_parameter_modulation_y}
\end{align}
where $\alpha_{\text{ref}}$ is the reference attenuation coefficient used during the initial optimization ($\alpha_{\text{ref}} = 0.5 \ \text{dB/cm/MHz}$ in this study). $y$ is kept 1.0 in this modulation. This linear scaling is an empirical approximation rather than a theoretically derived relation, and it is justified a posteriori by the fitting accuracy reported below.

\begin{figure}[t]
    \centering
    \begin{subfigure}[b]{0.9\linewidth}
        \centering
        \includegraphics[width=\linewidth]{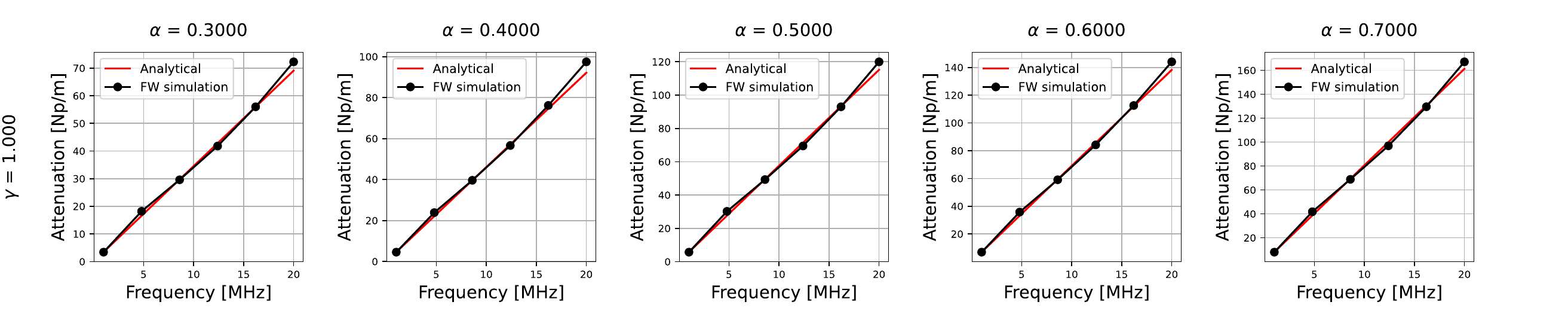}
        \caption{Attenuation fitting}
        \label{fig:attenuation_fitting}
    \end{subfigure}
    \begin{subfigure}[b]{0.9\linewidth}
        \centering
        \includegraphics[width=\linewidth]{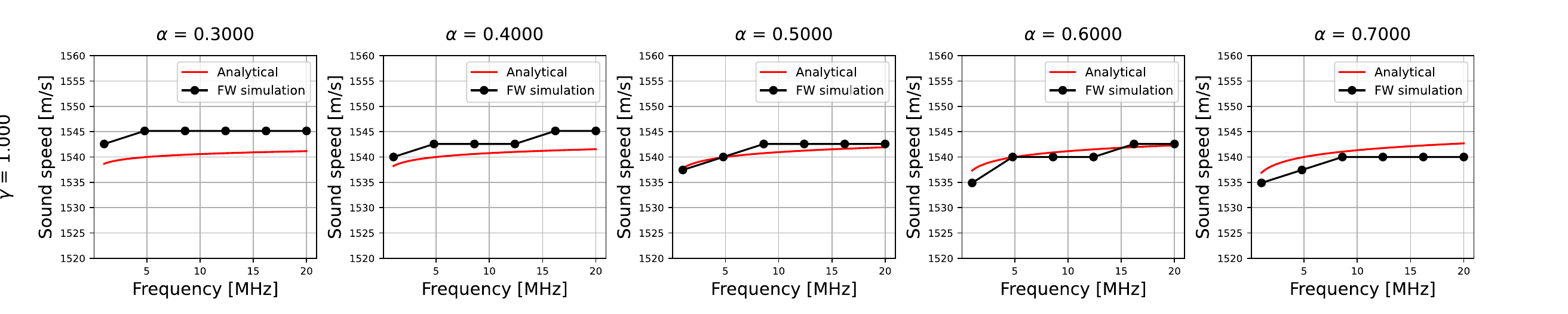}
        \caption{Dispersion fitting}
        \label{fig:phase_velocity_fitting}
    \end{subfigure}
    \caption{Comparison between simulated and target attenuation and phase velocity for various attenuation coefficients at a fixed power-law exponent $y=1.0$.}
    \label{fig:relaxation_fitting_results}
\end{figure}

To validate the accuracy of the multiple relaxation model in modeling arbitrary attenuation and dispersion, we performed a series of simulations to compare the predicted attenuation and phase velocity with the target power-law characteristics.
Figures \ref{fig:attenuation_fitting} and \ref{fig:phase_velocity_fitting} show the comparison between the simulated and target attenuation and phase velocity for various attenuation coefficients with power law exponent $y=1.0$.
The attenuation and phase velocity were calculated by measuring the amplitude of a Gaussian modulated sinusoidal pulse after propagating through a distance of $10 \lambda$ in a homogeneous medium.
The results demonstrate that the multiple relaxation model can accurately reproduce the desired attenuation characteristics over the frequency range of interest
(1-20 MHz).
It showed under $5\%$ error in attenuation for all tested power-law attenuation coefficients.
The phase velocity for $\alpha_0$ and $y=1.0$ closely matches the target values, while slight deviations are observed for different attenuation coefficients. This is attributed to the scaling approach used in Eq. (\ref{eq:relaxation_parameter_modulation_y}), which involves scaling only the $d_x^{\nu=1}$ parameters to achieve the desired attenuation characteristics.
However, the dispersion error remains within the error of $0.5\%$.

\subsection{Reflection coefficient and thickness of transition layer and C-PML} \label{sec:ref_coeff_vs_pml_thickness}
\begin{figure}[t]
    \centering
    \begin{minipage}[t]{0.45\textwidth}
        \centering
        \includegraphics[width=\linewidth]{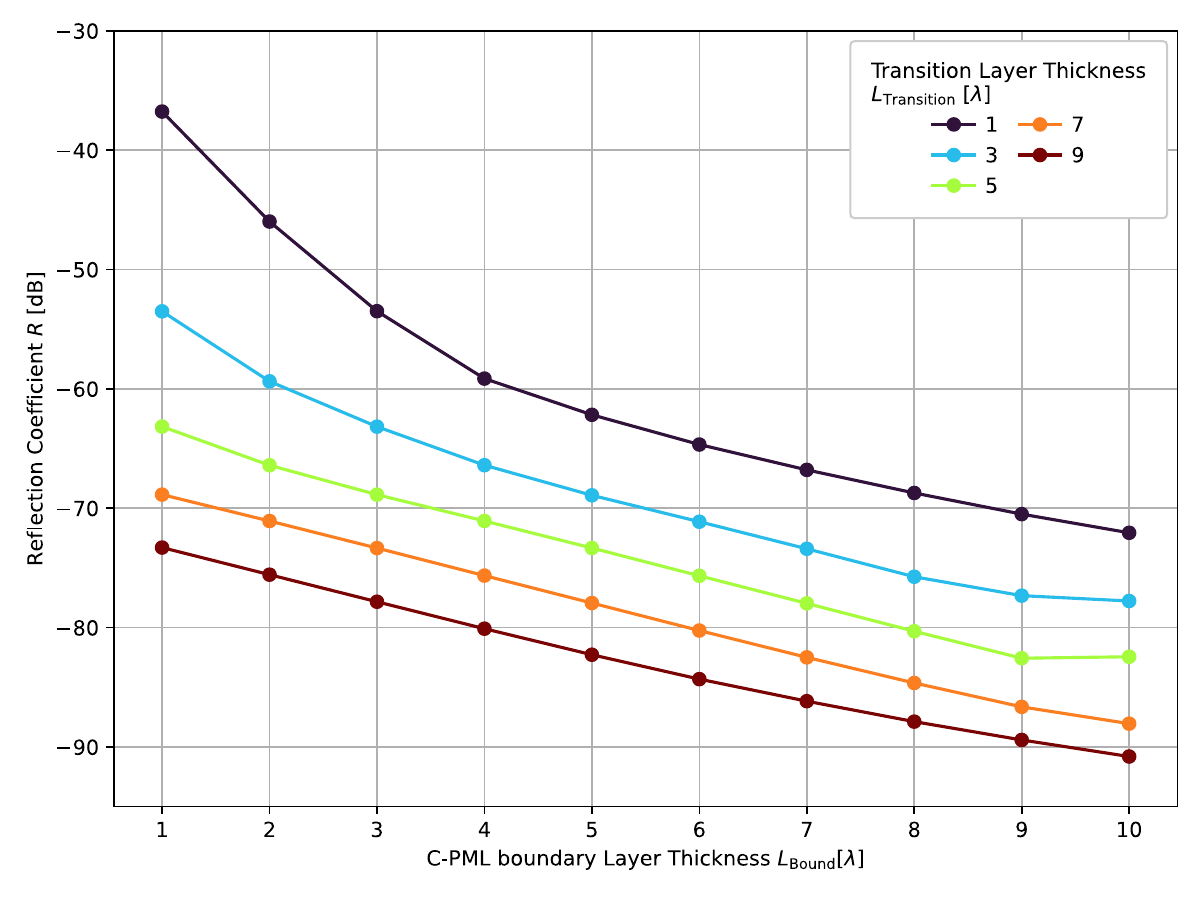}
        \subcaption{Reflection coefficient versus PML and transition layer thickness. The reflection coefficient improves as the thickness of the PML and transition layer increases.}
        \label{fig:ref_coeff_vs_pml_and_transition}
    \end{minipage}\hfill
    \begin{minipage}[t]{0.45\textwidth}
        \centering
        \includegraphics[width=\linewidth]{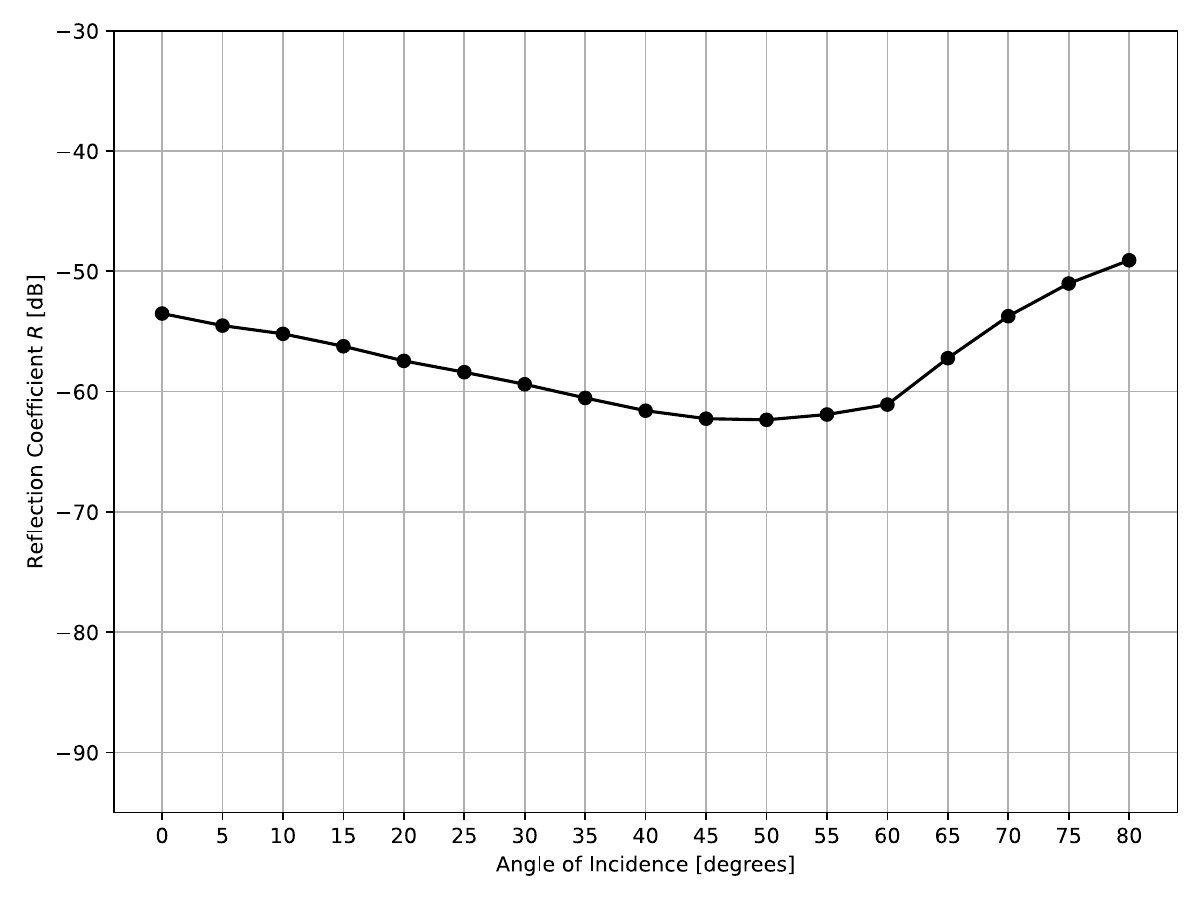}
        \subcaption{Reflection coefficient versus incident angle to the two stage PML with transition layer = 3 and PML = 1. The reflection coefficient is stable across different incident angles.}
        \label{fig:ref_coeff_vs_angle}
    \end{minipage}
    \caption{Reflection coefficient experiments for the two stage PML with transition layer.}
\end{figure}
We evaluate the performance of the proposed two-stage convolutional perfectly matched layer (C-PML) in minimizing reflections at the boundaries of the simulation domain.
The reflection coefficient is calculated by measuring the amplitude of a Gaussian modulated sinusoidal pulse before and after it interacts with the PML boundary.
The simulation is performed using a spatial resolution of 16 points per wavelength (PPW) and a Courant-Friedrichs-Lewy (CFL) number of 0.2.

Figure \ref{fig:ref_coeff_vs_pml_and_transition} shows the reflection coefficient when the PML layer thickness is varied for each transition-layer thickness. It is observed that the reflection coefficient improves as the thickness of the PML and transition layer increases.
As a result, if the target reflection coefficient is $-50 \ \text{dB}$, $L_{\text{Transition}} = 3\lambda$ and $L_{\text{Bound}} = 1\lambda$ are preferable considering the trade-off between computational cost and performance. It adds up to 4$\lambda$ grid points in total in the PML region.
If the target reflection coefficient is $-60 \ \text{dB}$, $L_{\text{Transition}} = 5\lambda$ and $L_{\text{Bound}} = 1\lambda$ are preferable.
Based on these results, we use $L_{\text{Transition}} = 3\lambda$ and $L_{\text{Bound}} = 1\lambda$ in subsequent experiments considering the trade-off between computational cost and performance.
\subsection{Reflection coefficient and incident angle} \label{sec:ref_coeff_vs_angle}

Similar to the previous subsection, we evaluate the reflection coefficient of the proposed two-stage convolutional perfectly matched layer (C-PML) as the incident angle of the incoming wave varies.
The simulation setup is similar to the previous subsection, but the incident angle of the Gaussian modulated sinusoidal pulse is varied from normal incidence (0 degrees) to almost grazing incidence (80 degrees).
Figure \ref{fig:ref_coeff_vs_angle} illustrates the reflection coefficient as the incident angle to the two-stage PML varies. The two-stage PML maintains a reflection coefficient below about $-49$~dB across the full range of tested incidence angles, with a minimum close to $-62$~dB near $45$ degrees. This angular dependence follows from the directional implementation of the PML, in which the complex coordinate stretching is applied independently along the $x$, $y$, and $z$ directions. Near $0$ and $90$ degrees the wave propagates essentially along a single axis and is absorbed mainly by one directional PML, whereas near $45$ degrees it has comparable components along both axes and is absorbed by the $x$- and $y$-direction PMLs together, which yields the lowest reflection.

The increase in reflection toward the largest incidence angles is consistent with the reduced absorption of absorbing boundaries near grazing incidence. \citet{Gao2017-tp}, in their Fig.~3, show that for one-way wave-equation absorbing boundary conditions the reflection coefficient rises steeply toward grazing incidence, and their large-scale numerical experiments indicate that nearly grazing incident waves are the most difficult for the PML to absorb. The reflection coefficient of the two-stage PML nonetheless remains low across all tested angles, which indicates that it is robust to the incidence angle.

\subsection{Spatial and temporal discretization effects}
\begin{figure}[t]
    \centering
    \includegraphics[width=0.8\linewidth]{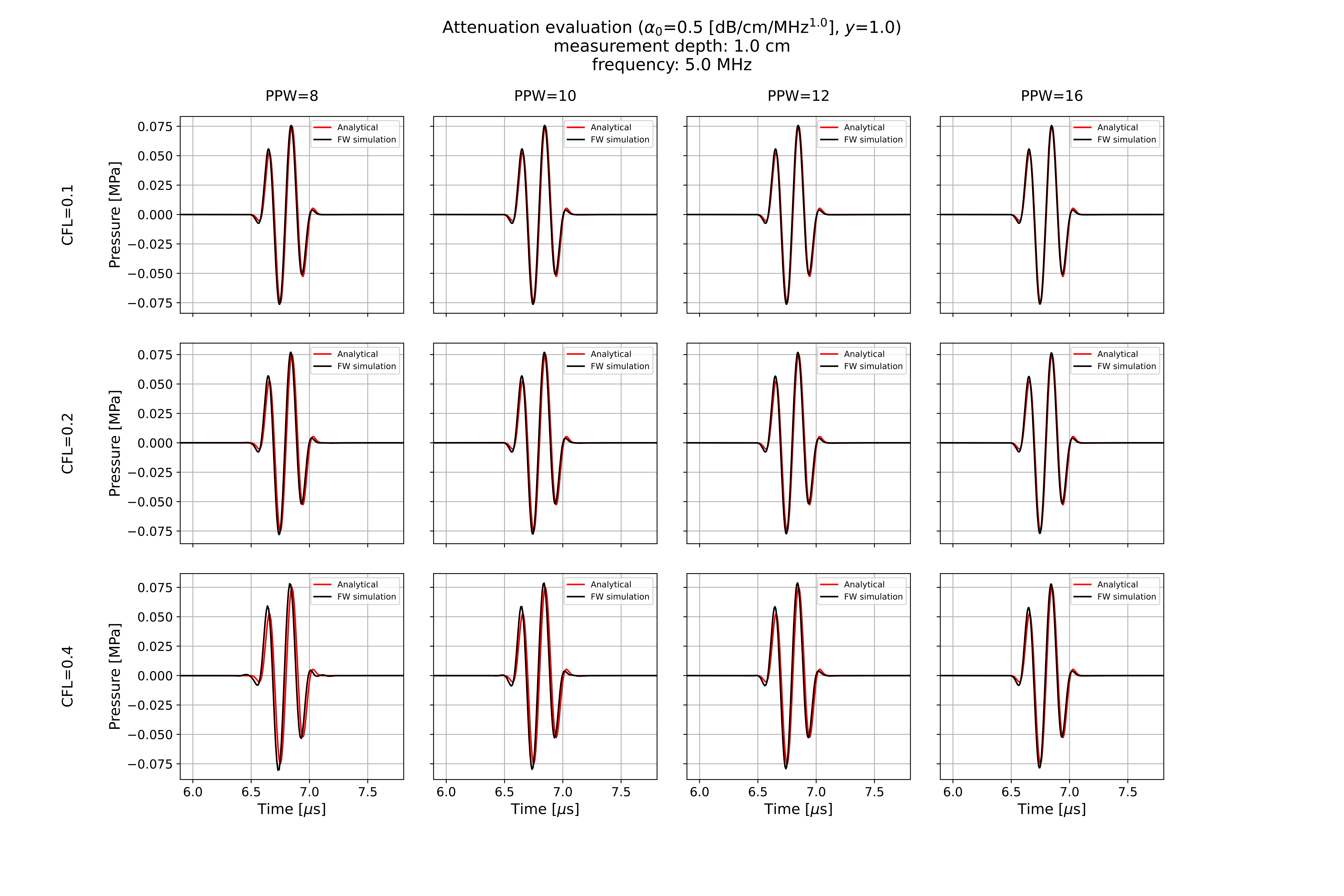}
    \caption{Pulse shape when the spatial and temporal discretization are varied. The spatial discretization is varied by changing the points per wavelength (PPW), and the temporal discretization is varied by changing the Courant-Friedrichs-Lewy (CFL) number. A $5 \text{MHz}$ Gaussian modulated sinusoidal pulse is used as the input pulse. The figure shows the pulse shape after propagating through a distance of $32$ wavelengths. The pulse shape is compared against the angular spectrum method. While it shows the accuracy of wave location decreases as PPW decreases and CFL increases, the pulse shape is well-preserved even when PPW is 8 and CFL is 0.4.}
    \label{fig:pulse_shape}
\end{figure}
Figure \ref{fig:pulse_shape} shows the pulse shapes after propagating through an attenuating medium ($\alpha_0=0.5$, $y=1.0$) for a distance of 32 wavelengths. Each subplot shows the pulse shape when the spatial and temporal discretization are varied.
The spatial discretization is varied by changing the points per wavelength (PPW), and the temporal discretization is varied by changing the Courant-Friedrichs-Lewy (CFL) number.
A $5 \text{MHz}$ Gaussian modulated sinusoidal pulse is used as the input pulse.
The pulse shape is compared against the angular spectrum method, which is a highly accurate method for simulating wave propagation in homogeneous media.
While it shows the accuracy of wave location decreases as PPW decreases and CFL increases, the pulse shape is well-preserved even when PPW is 8 and CFL is 0.4.

\subsection{Nonlinearity evaluation}
\begin{figure}[t]
    \centering
    \begin{minipage}[t]{0.45\textwidth}
        \includegraphics[width=1.0\linewidth]{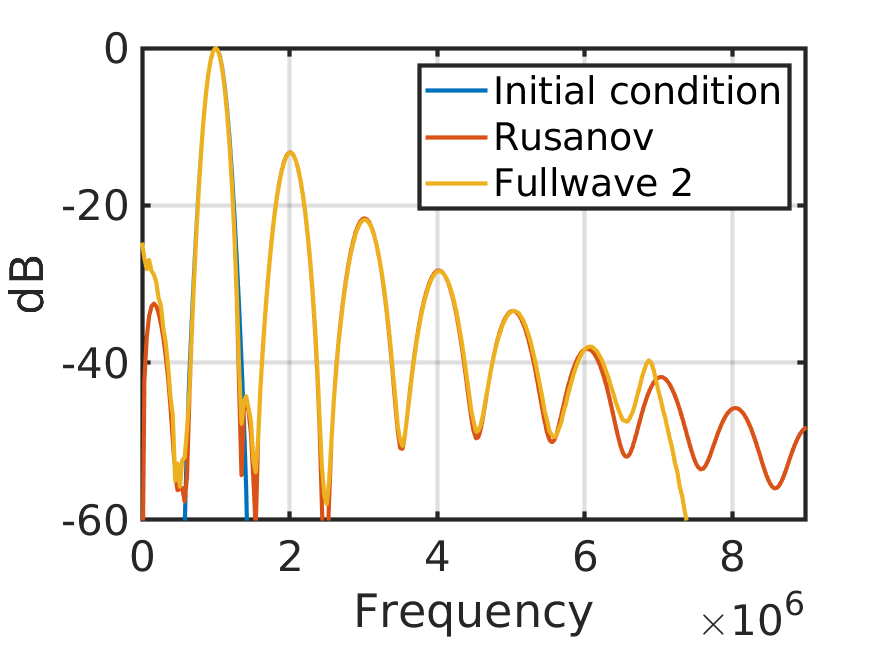}
        \subcaption{Frequency domain comparison}
        \label{fig:nln_fit}
    \end{minipage}
    \begin{minipage}[t]{0.45\textwidth}
        \includegraphics[width=1.0\linewidth]{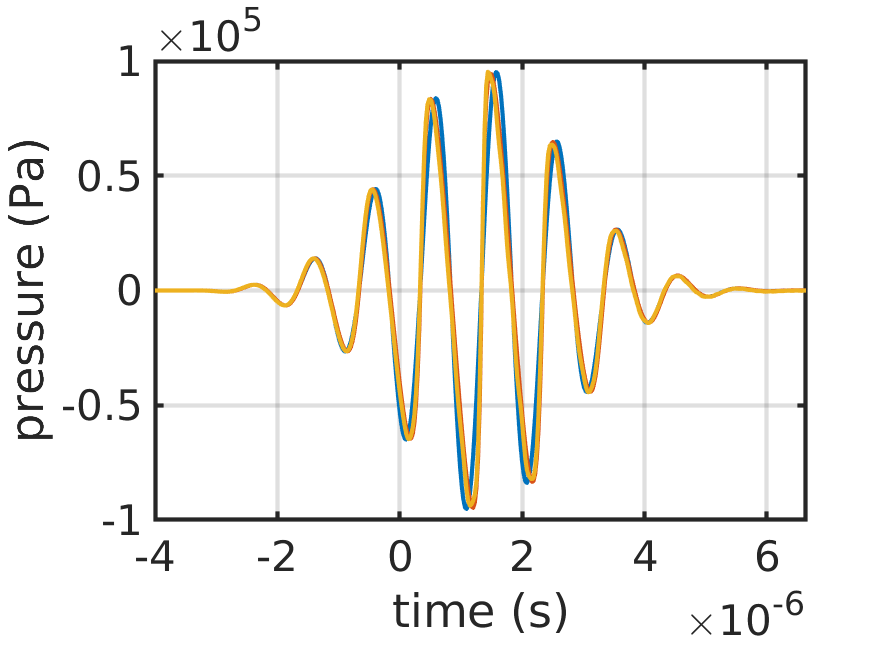}
        \subcaption{Time domain comparison}
        \label{fig:nln_fit_waveform}
    \end{minipage}
    \caption{Comparison of nonlinearity described by Rusanov 1D solution to Burgers equation using 150 points per wavelength and Fullwave solution with a 15 points per wavelength grid. Fullwave solution is accurate up to the $7^{\text{th}}$ harmonic, indicating that the spectral support fails just before Nyquist, i.e., the fundamental sampling limit for this grid size, which occurs at 7.5 $f_0$. The initial-condition curve is the undistorted source pulse at the input plane, a Gaussian-modulated sinusoid with amplitude $p_0 = 100$~kPa, center frequency $f_0 = 1$~MHz, and 4 cycles, shown for reference against the steepened solutions. The horizontal axis in (b) is time referenced to the center of the transmitted pulse, so samples preceding the pulse center are negative. The comparison is performed at a dimensionless nonlinearity parameter $\sigma \approx 0.7$, in the preshock regime ($\sigma < 1$).}
\end{figure}

To evaluate the nonlinearity implementation in Fullwave~2, we compared the simulation results with a 1D Rusanov solution to the inviscid quadratic Burgers equation.
A 2D plane wave was transmitted in a material with no attenuation/dispersion and compared to 1D solutions of the inviscid
quadratic Burgers equation solved with a Rusanov scheme. The reference Rusanov solutions were implemented on a grid
that was 10 times finer (150 points per wavelength) than the Fullwave solver (15 points per wavelength). A close match is obtained both in the frequency (Fig. \ref{fig:nln_fit}) and time domains (Fig. \ref{fig:nln_fit_waveform}). The frequency domain error shows that the Fullwave solution is accurate up to the $7^{\text{th}}$ harmonic, indicating that the spectral support fails just before Nyquist, i.e., the fundamental sampling limit for this grid size, which occurs at 7.5 $f_0$.

The strength of the nonlinear distortion is characterized by the dimensionless parameter $\sigma = \beta \varepsilon k x$, a standard measure in nonlinear acoustics \citep[see][Eq.~4.23]{HamiltonBlackstock2024-os}, where $\beta = 1 + B/2A$ is the coefficient of nonlinearity, $\varepsilon$ the acoustic Mach number, $k$ the wavenumber, and $x$ the propagation distance. It expresses the propagation distance in units of the shock-formation distance, so that $\sigma = 1$ marks the onset of shock formation. The comparison in Fig.~\ref{fig:nln_fit} is performed at $\sigma \approx 0.7$, in the preshock regime ($\sigma < 1$) where the wave has steepened but not yet formed a shock.

\subsection{Computational performance}
\begin{figure}[t]
    \centering
    \begin{minipage}[t]{0.4\textwidth}
        \includegraphics[width=1.0\linewidth]{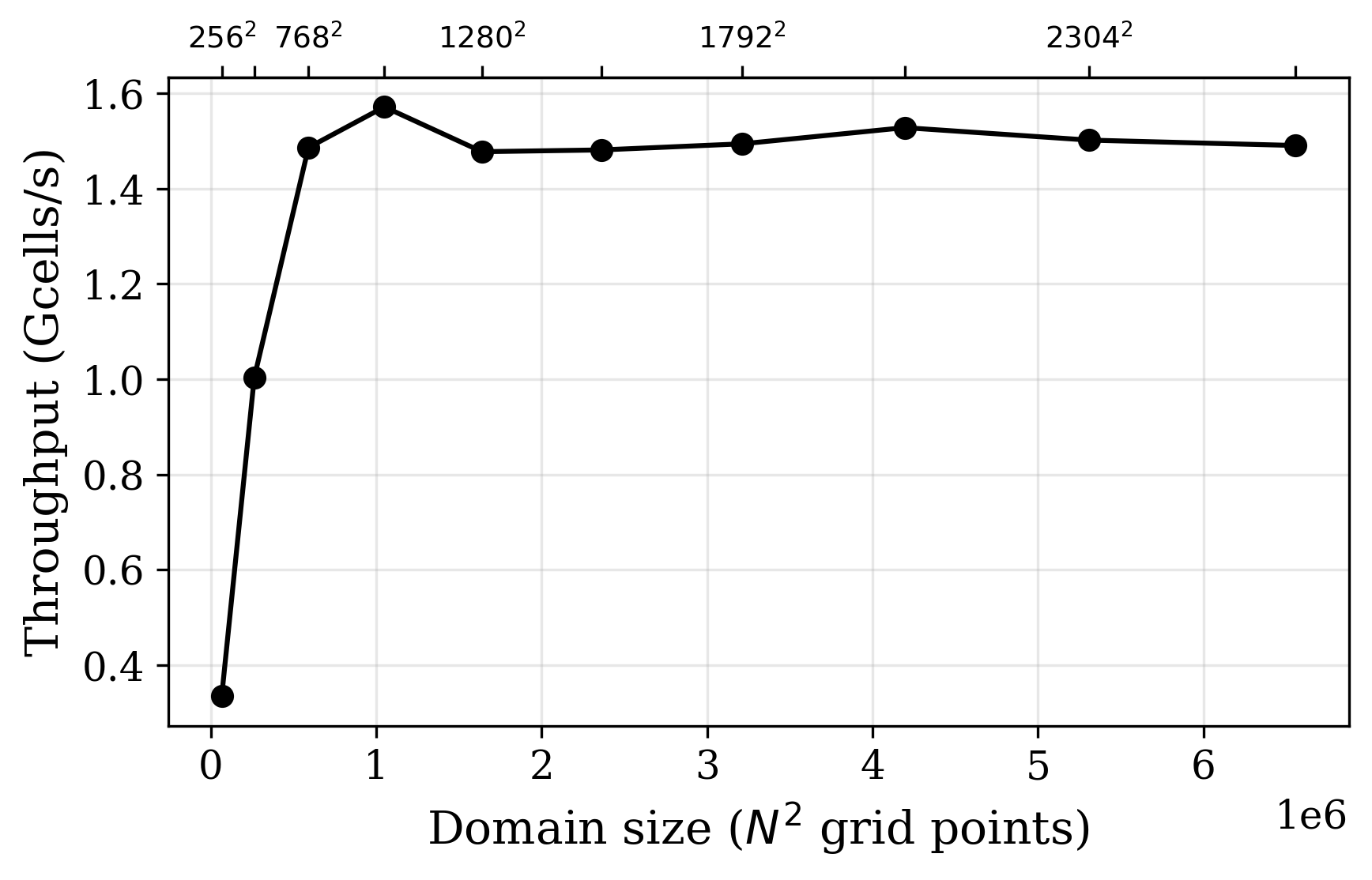}
        \subcaption{2D Throughput vs Number of grid points}
        \label{fig:single_gpu_throughput}
    \end{minipage}
    \begin{minipage}[t]{0.4\textwidth}
        \includegraphics[width=1.0\linewidth]{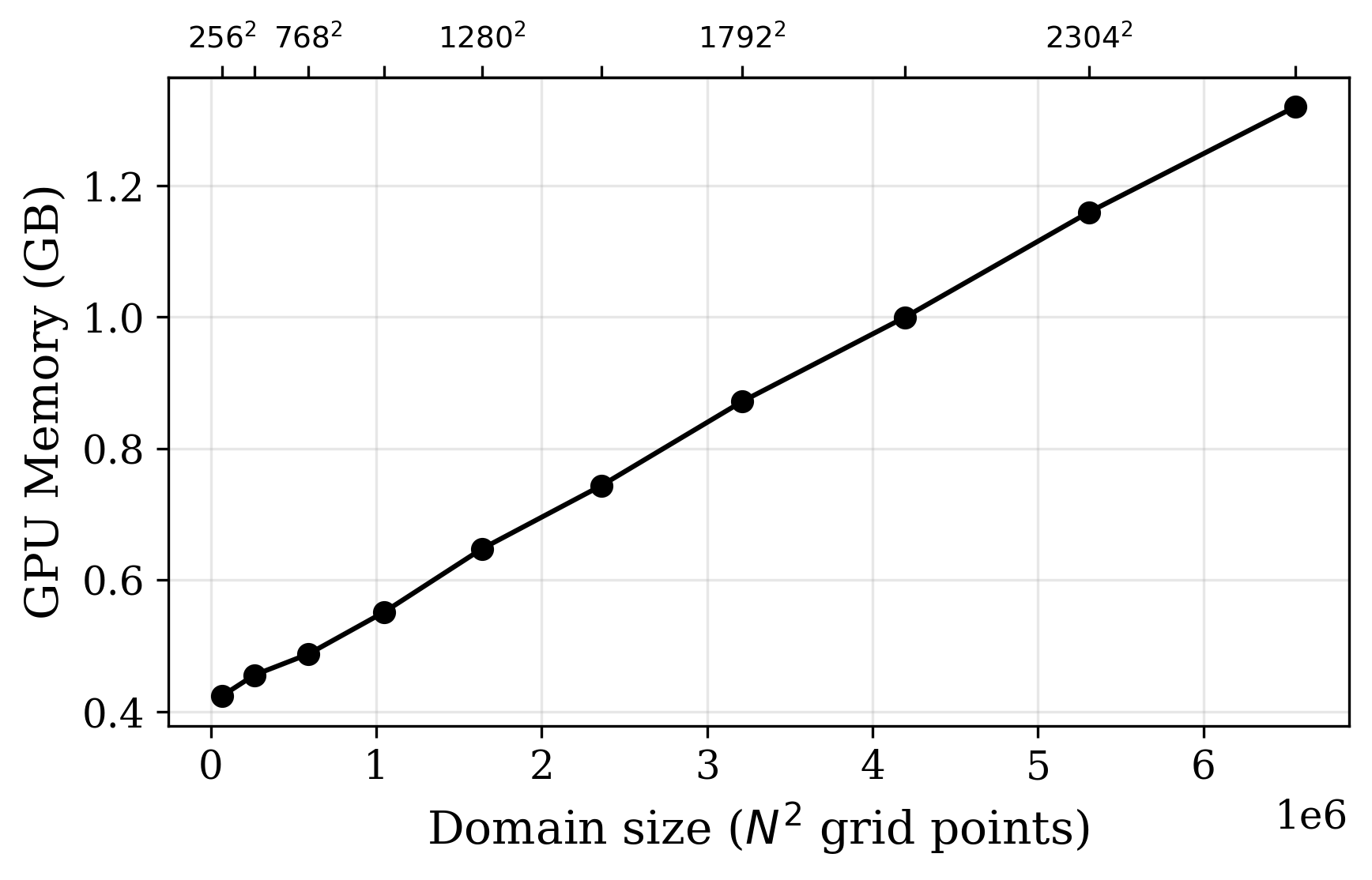}
        \subcaption{2D Memory usage vs Number of grid points}
        \label{fig:single_gpu_memory_usage}
    \end{minipage}
    \begin{minipage}[t]{0.4\textwidth}
        \includegraphics[width=1.0\linewidth]{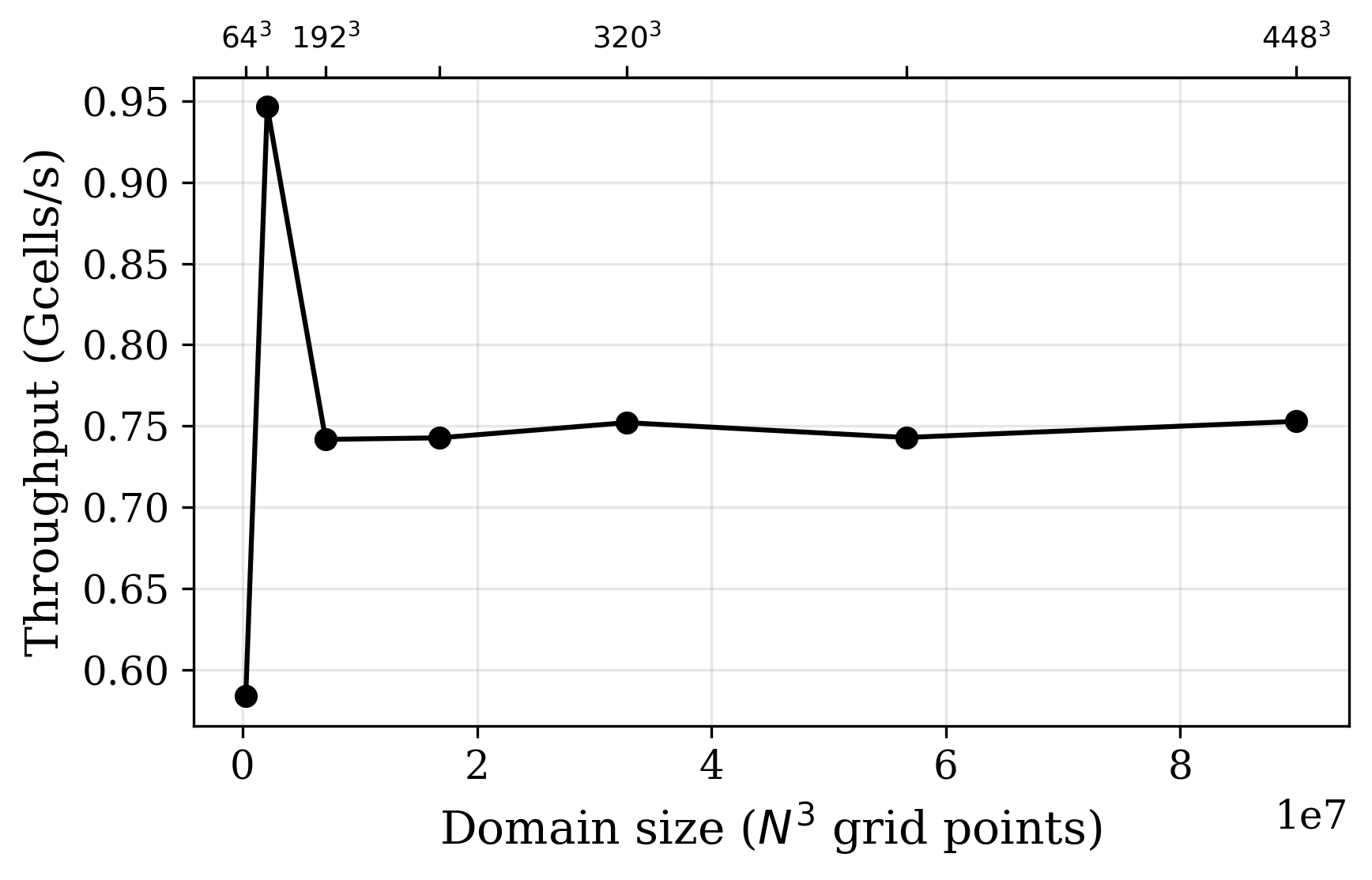}
        \subcaption{3D Throughput vs Number of grid points}
        \label{fig:single_gpu_throughput_3d}
    \end{minipage}
    \begin{minipage}[t]{0.4\textwidth}
        \includegraphics[width=1.0\linewidth]{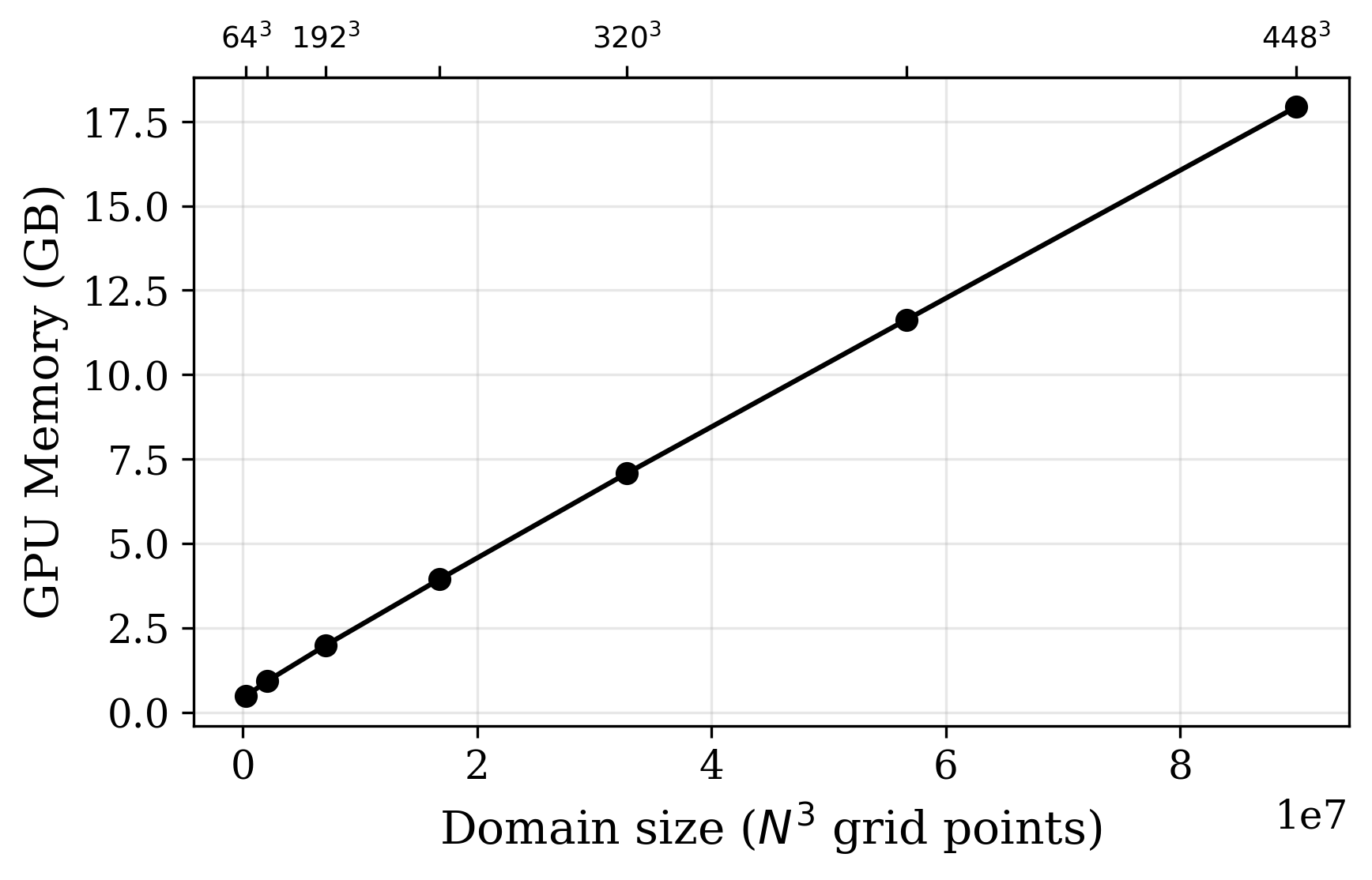}
        \subcaption{3D Memory usage vs Number of grid points}
        \label{fig:single_gpu_memory_usage_3d}
    \end{minipage}
    \caption{Benchmark results of Fullwave~2 on a single NVIDIA RTX 4090 (24 GB, CUDA 12.6, Python 3.12). (a) Throughput in cells per second (GCells/sec) as the number of grid points increases in 2D. (b) Memory usage in GB as the number of grid points increases in 2D. (c) Throughput in cells per second (GCells/sec) as the number of grid points increases in 3D. (d) Memory usage in GB as the number of grid points increases in 3D.}
\end{figure}

We benchmarked Fullwave~2 on a single NVIDIA RTX 4090 (24 GB, CUDA 12.6, Python 3.12).
Figure \ref{fig:single_gpu_throughput} and \ref{fig:single_gpu_memory_usage} show the throughput in G cells per second (GCells/sec) and memory usage in GB as the number of grid points increases in 2D, respectively.
Figures \ref{fig:single_gpu_throughput_3d} and \ref{fig:single_gpu_memory_usage_3d} show the throughput and memory usage in 3D, respectively.
Figure \ref{fig:single_gpu_throughput} and \ref{fig:single_gpu_throughput_3d} show that the throughput increases with the number of grid points until it saturates at a certain point.
This saturation point is determined by the GPU's computational capabilities and memory bandwidth.
At 12 PPW and CFL = 0.4, throughput is $\sim\!1.5$ GCells/sec in 2D and $\sim\!0.75$ GCells/sec in 3D with $N=2$ relaxations.
Figure \ref{fig:single_gpu_memory_usage} and \ref{fig:single_gpu_memory_usage_3d} show that the memory usage increases linearly with the number of grid points.

For practical reference in 2D simulations, a $1024^2$ grid with $N=2$ fits in $\sim$0.55 GB and runs $\sim$0.667 seconds for 1{,}000 steps; a $2048^2$ grid runs $\sim$2.74 seconds for 1{,}000 steps with 1.0 GB memory usage.
For 3D simulations, a $256^3$ grid with $N=2$ fits in $\sim$3.9 GB and runs $\sim$22.6 seconds for 1{,}000 steps.

\section{Discussion}

We have presented a flexible and efficient formulation to model arbitrary power-law attenuation and dispersion in biological tissues using the multiple relaxation model within a high-order finite difference framework on a staggered grid.
By optimizing the relaxation parameters, we can closely approximate the desired power-law behavior over a specified frequency range.
The proposed approach effectively captures the desired attenuation and dispersion characteristics, as validated through numerical simulations comparing the predicted and target behaviors.

\subsection{Applications}

The proposed multiple relaxation model can be applied to various ultrasound simulation scenarios, including medical imaging, therapeutic ultrasound and transcranial ultrasound.
The ability to represent arbitrary power-law attenuation allows for more accurate simulations of different tissue types, leading to improved image quality and diagnostic accuracy.

Supplementary Material \ref{appendix:abdominal_wall_imaging} demonstrates the 2D application of the proposed method in simulating B-mode ultrasound imaging of an abdominal wall phantom with heterogeneous attenuation properties.
A convex transducer array (C5-2V) with 128 elements operating at a center frequency of 3.7 MHz was used to simulate the imaging process. We employed a full synthetic aperture imaging technique, where each element sequentially transmits a pulse while all elements receive the echoes.
It shows the capability of the proposed method to model complex attenuation, aberration, and reverberation effects in abdominal wall imaging.
The accurate representation of these effects is essential for developing and evaluating advanced imaging techniques, such as adaptive beamforming, aberration correction algorithms and deep learning-based image enhancement methods, which can significantly improve image quality in challenging clinical scenarios.

Supplementary Material \ref{appendix:rat_skull_imaging} illustrates the 3D application of the proposed method in simulating transcranial ultrasound imaging of a rat brain through the skull bone with heterogeneous attenuation properties.
A 2D matrix transducer array with 1024 elements operating at a center frequency of 7.81 MHz was used to simulate a single plane wave transmission.
The ability to model the complex attenuation and reverberation characteristics of the skull bone is essential for improving the image quality in transcranial ultrasound applications such as brain imaging and neuromodulation \citep{Soulioti2025-pa}.

\subsection{Fitting accuracy}
We have obtained the relaxation parameters through a coarse-to-fine grid search to minimize the fitting error between the target and modeled attenuation and phase velocity over the frequency range of interest. The fitting accuracy offers a sufficient accuracy for practical ultrasound simulations.
We achieved a fitting error of less than 5$\%$ over the frequency range of 1 to 20 MHz using two relaxation mechanisms.
This level of accuracy is adequate since the experimental measurements of tissue attenuation and dispersion often exhibit higher variability than this fitting error.

\subsection{Advantages of the multiple relaxation model}
The finite difference implementation of the multiple relaxation model offers several advantages for modeling arbitrary attenuation and dispersion in ultrasound simulations. First, the multiple relaxation model provides a capability to approximate a wide range of frequency-dependent attenuation behaviors by adjusting the relaxation parameters. This flexibility allows for accurate representation of the complex acoustic properties of biological tissues, which often exhibit non-standard attenuation characteristics that deviate from simple power-law models.

Second, the uniform framework of the extended C-PML for the multiple relaxation model allows for seamless integration of attenuation and dispersion effects into the wave propagation simulations while easily implementing perfectly matched layers (PML) to minimize reflections at the boundaries. This integration is crucial for accurately simulating wave propagation in heterogeneous media, such as human tissues, where boundary reflections can significantly affect the results.

Third, finite difference implementations, such as the one presented here, are suitable for large-scale simulations due to their spatial and temporal locality. This locality allows for efficient memory usage and parallelization, making the method well-suited for high-performance computing environments such as multiple GPUs or distributed computing clusters.

Finally, the use of a staggered grid in the finite difference scheme enhances numerical stability and accuracy, particularly for wave propagation problems. The staggered grid allows for better representation of wavefields and reduces numerical dispersion, which is crucial for accurately capturing the effects of attenuation and dispersion over long propagation distances. This implementation makes it possible to model subresolution scatterers without introducing significant numerical artifacts.

\subsection{Relation between reflection coefficient, PML and transition layer thickness}

The proposed two-stage PML method effectively minimizes reflections at the boundaries, ensuring accurate wave propagation in simulations even in the presence of multiple relaxation mechanisms.
The reflection coefficient decreases as the PML thickness increases, as shown in Fig. \ref{fig:ref_coeff_vs_pml_and_transition}.
Furthermore, introducing a transition layer before the PML further reduces the reflection coefficient.
As a result, if the target reflection coefficient is -50 dB, transition layer = 3 and PML = 1 are preferable considering the trade-off between computational cost and performance.

\subsection{Spatial and temporal discretization effects}

Based on the results of the spatial and temporal discretization effects on attenuation modeling, the discretization effects are negligible when using 12 points per wavelength (PPW) and a CFL number of 0.4 or less. Furthermore, the pulse shape is well-preserved even with 8 PPW and a CFL number of 0.4. This numerical stability is inherent to Tan et al.'s staggered grid finite difference method \citep{Tan2014-qt}. Reducing the PPW to 8 and the CFL number to 0.2 still maintains good pulse shape preservation, although there is a slight increase in attenuation modeling error. The 8 PPW with CFL number of 0.2 is a very efficient setting that can reduce the GPU memory usage to one eighth compared to the 16 PPW in 3D simulation. This setting is useful for reduced order large scale simulation.

\subsection{Limitations}

While the proposed method effectively models arbitrary power-law attenuation and dispersion using the multiple relaxation model, several limitations should be acknowledged.

First, the grid-search based optimization approach for the relaxation parameters is computationally intensive and may not guarantee finding the global optimum. The coarse-to-fine search for one power law parameter such as $\alpha_0=0.5, y=1.0$ requires exploring a large number of parameter combinations.
For instance, in this study, we performed a grid search with 5 spaced values in each dimension for 10 relaxation parameters in the coarse search, followed by 4 additional finer searches with 5 spaced values in each dimension. This process involves evaluating $5 \times 5^{10} \approx 49 ~  \text{million}$ parameter combinations. The coarse-to-fine grid search took about 20 minutes on a 24-core Intel Core i9-13900K server. Additionally, the grid search intervals were manually adjusted based on the previous search results to ensure convergence toward the optimal parameters, which is not suitable for fully automated parameter fitting.
While this process provides reasonable approximations for various attenuation parameters through scaling (Eq.~(\ref{eq:relaxation_parameter_modulation_y})), the optimization time remains a bottleneck if we aim to fit a wide range of tissue types with different attenuation characteristics.

Future work will focus on automating the optimization process to enable rapid parameter fitting across diverse tissue types and frequency ranges. Two promising directions are: (i) derivative-free optimization algorithms such as COBYLA, which can efficiently navigate the high-dimensional parameter space without requiring gradient information; and (ii) backpropagation-based gradient methods, potentially reducing optimization time by orders of magnitude. Additionally, we will investigate whether a small database of pre-optimized parameters for common tissue types (liver, muscle, fat) can serve as initialization seeds, further accelerating the fitting process for new attenuation characteristics.

Second, the proposed model requires additional computational resources due to the inclusion of multiple relaxation mechanisms. Although the increase in memory usage and computation time is manageable for a moderate number of relaxation processes, it may become significant for very high-fidelity simulations requiring numerous relaxation terms.

Third, the current implementation of the convolutional perfectly matched layer (C-PML) assumes a polynomial increase in the attenuation constant within the PML region. While this approach effectively minimizes reflections, further investigation into alternative PML formulations for the multiple relaxation model could yield improved performance.

\subsection{Extension to full waveform inversion}

Beyond the forward modeling presented here, the multiple-relaxation formulation provides a natural basis for the inversion of tissue parameters by full waveform inversion. The gradient of a data misfit with respect to the medium and relaxation parameters can be obtained either by an adjoint method or by automatic differentiation, the latter often referred to as the backpropagation method. A consideration specific to this formulation is that the attenuation modeled by the multiple-relaxation mechanism is not time-reversible, so the adjoint of the relaxation operators must be derived explicitly rather than obtained by a simple time reversal of the forward operators. Attenuation-aware inversion built on this formulation is a promising direction for future work.

\section{Conclusion}

We presented Fullwave~2, a nonlinear attenuating wave equation formulation that preserves the staggered-grid d'Alembertian structure while incorporating multiple relaxation mechanisms and a two-stage C-PML. The model achieves $<5\%$ attenuation error and $<0.5\%$ phase-velocity error over 1--20 MHz, and its PML yields reflection coefficients below $-49\ \mathrm{dB}$ with a $4\lambda$ boundary thickness. Nonlinear propagation matches a Burgers reference up to the $7^{\text{th}}$ harmonic, and ultrasound-scale demonstrations in 2D and 3D show accurate long-range propagation in heterogeneous media.
The multiple relaxation model's flexibility allows for an accurate representation of various tissue types. This facilitates the execution of realistic simulations for medical imaging, therapeutic ultrasound applications, and transcranial ultrasound. Accurate modeling of attenuation is also essential for creating training datasets for machine learning applications and for mitigating the difference between the simulated and real domains.

Future work includes automating relaxation-parameter fitting, exploring alternative PML formulations tailored to multiple relaxations, and extending the solver to richer tissue models and clinical imaging scenarios.

\ifarxiv
    \section*{Acknowledgments}
    Funding provided by NIH R01EB029419 and R01EB036295. We would like to thank the University of North Carolina at Chapel Hill and the Research Computing group for providing computational resources and support that have contributed to these research results.
\else
    \ack{Funding provided by NIH R01EB029419 and R01EB036295. We would like to thank the University of North Carolina at Chapel Hill and the Research Computing group for providing computational resources and support that have contributed to these research results.}
\fi

\ifarxiv
    \section*{Data availability}
    An open-source implementation of the solver is available as part of the companion Fullwave 2.5 release at \url{https://github.com/pinton-lab/fullwave25}, which uses the same finite-difference time-domain and C-PML formulation.
\else
    \data{An open-source implementation of the solver is available as part of the companion Fullwave 2.5 release at \url{https://github.com/pinton-lab/fullwave25}, which uses the same finite-difference time-domain and C-PML formulation.}
\fi

\ifarxiv\else
    \suppdata{
\begin{appendices}
    \section{Derivation of dispersion relation for the multiple relaxation model} \label{appendix:dispersion_relation}
    The derivation of the dispersion relation Eq.~(\ref{eq:dispersion_relations}) is provided below.
    Assuming that the governing equation is 1-dimensional plane wave, we obtain the solution $p(x, t) = p_0 e^{i (\omega t - kx)}$ and $v(x, t) = v_0 e^{i (\omega t - kx)}$. Taking the Fourier transform of Eq.~(\ref{eq:convolution_kernel_x}) yields,
    \begin{align}
        F\left[\zeta^\nu_{x_1}(t)\right] = -\cfrac{d^\nu_{x1}}{\kappa^2_{x1}} \left( \cfrac{1}{d_{x1}^\nu / \kappa_{x1} + \alpha_{x1}^{\nu} + i \omega} \right) = \tilde{\zeta}(\omega)
    \end{align}
    Considering convolution relationship in Fourier space and plane wave solution, the convolution term inside the summation in Eq. (\ref{eq:relaxation1}) for pressure will be
    \begin{align}
        F\left[\zeta(t) \ast p(x, t)\right] & = \tilde{\zeta}(\omega) \cdot F\left[p(x, t)\right]                                                                  \\
                                                  & = \tilde{\zeta}(\omega) \cdot \left[2 \pi p_0 e^{ikx} \delta(\omega - \omega_0)\right] \label{eq:fourier_plane_wave}
    \end{align}
    Taking an inverse Fourier transform of Eq.~(\ref{eq:fourier_plane_wave}) gives
    \begin{align}
        F^{-1}\left[\tilde{\zeta}(\omega) \cdot \left[2 \pi p_0 e^{ikx} \delta(\omega - \omega_0)\right]\right] & = \cfrac{1}{2 \pi} \int^\infty_{-\infty} \tilde{\zeta}(\omega) \cdot 2 \pi e ^ {ikx} \cdot e^{-i \omega t} \delta(\omega - \omega_0) d \omega             \\
                                                                                                                & = p_0 e^{ikx} \tilde{\zeta} (\omega_0) \int^{\infty}_{-\infty} \delta(\omega - \omega_0) e^{-i \omega t} d\omega                                          \\
                                                                                                                & =                                                                                 \tilde{\zeta}(\omega_0) \cdot p_0 e^{i (kx - \omega_0 t)}               \\
                                                                                                                & =                                                                                 \tilde{\zeta}(\omega_0) \cdot p(x, t) \label{eq:convolution_plane_wave}
    \end{align}
    Therefore, using Eq. (\ref{eq:gamma1}), Eq. (\ref{eq:convolution_plane_wave}), and Eq. (\ref{eq:relaxation1}),
    \begin{align}
        \cfrac{\partial p}{\partial \tilde{x}_1} & = \cfrac{1}{\kappa_{x_1}} \cdot \cfrac{\partial p}{\partial x} + \sum_{\nu=1}^{N} \zeta_{x_1}^\nu \ast \cfrac{\partial p}{\partial x} \\
                                                 & =  \cfrac{1}{\kappa_{x_1}} \cdot \cfrac{\partial p}{\partial x} + \sum^{N}_{\nu=1} \tilde{\zeta}(\omega) (p ik)                       \\
                                                 & = \cfrac{1}{\kappa_{x_1}} \cdot ikp - \gamma_1 \cdot ikp                                                                              \\
                                                 & = ikp \left[\cfrac{1}{\kappa_{x_1}} - \gamma_1\right] \label{eq:relaxation_derivative}
    \end{align}
    substituting Eq.~(\ref{eq:relaxation_derivative}) to (\ref{eq:wave1}), we obtain
    \begin{align}
        ik \left(\cfrac{1}{\kappa_{x_1}} - \gamma_1\right) p - i\omega \rho v = 0 \\
        \therefore \ v = \cfrac{k}{\omega \rho} \left(\cfrac{1}{\kappa_{x_1}} - \gamma_1\right) p \label{eq:wave1_velocity}
    \end{align}
    substituting Eq.~(\ref{eq:relaxation_derivative}) to (\ref{eq:wave2}), where $K = 1/(\rho c^2)$ is the medium compressibility, we obtain
    \begin{align}
        \nabla_2 \cdot \matr{v} & = \left(\cfrac{1}{\kappa_{x_2}} - \gamma_2\right) \cdot ik\matr{v}                                 \\
        \therefore v            & = \cfrac{\omega K p}{k \left(\cfrac{1}{\kappa_{x_2}} - \gamma_2 \right)} \label{eq:wave2_velocity}
    \end{align}
    Equating Eq. (\ref{eq:wave1_velocity}) and Eq. (\ref{eq:wave2_velocity}), we obtain the dispersion relation Eq. (\ref{eq:dispersion_relations}).

    \begin{align}
        k^2 \left(\cfrac{1}{\kappa_{x_1}} - \gamma_1\right) \left(\cfrac{1}{\kappa_{x_2}} - \gamma_2\right)  = \omega ^ 2 \rho K
        = \cfrac{\omega ^ 2}{c ^ 2} \\
        \therefore k = \cfrac{\omega}{c} \left( \cfrac{1}{\kappa_{x1} \kappa_{x2}} - \cfrac{\gamma_1}{\kappa_{x2}} - \cfrac{\gamma_2}{\kappa_{x1}} + \gamma_1 \gamma_2 \right) ^ {-\frac{1}{2}}
    \end{align}
    \begin{figure}[b]
        \centering
        \begin{minipage}[t]{0.28\textwidth}
            \centering
            \includegraphics[width=1.0\linewidth]{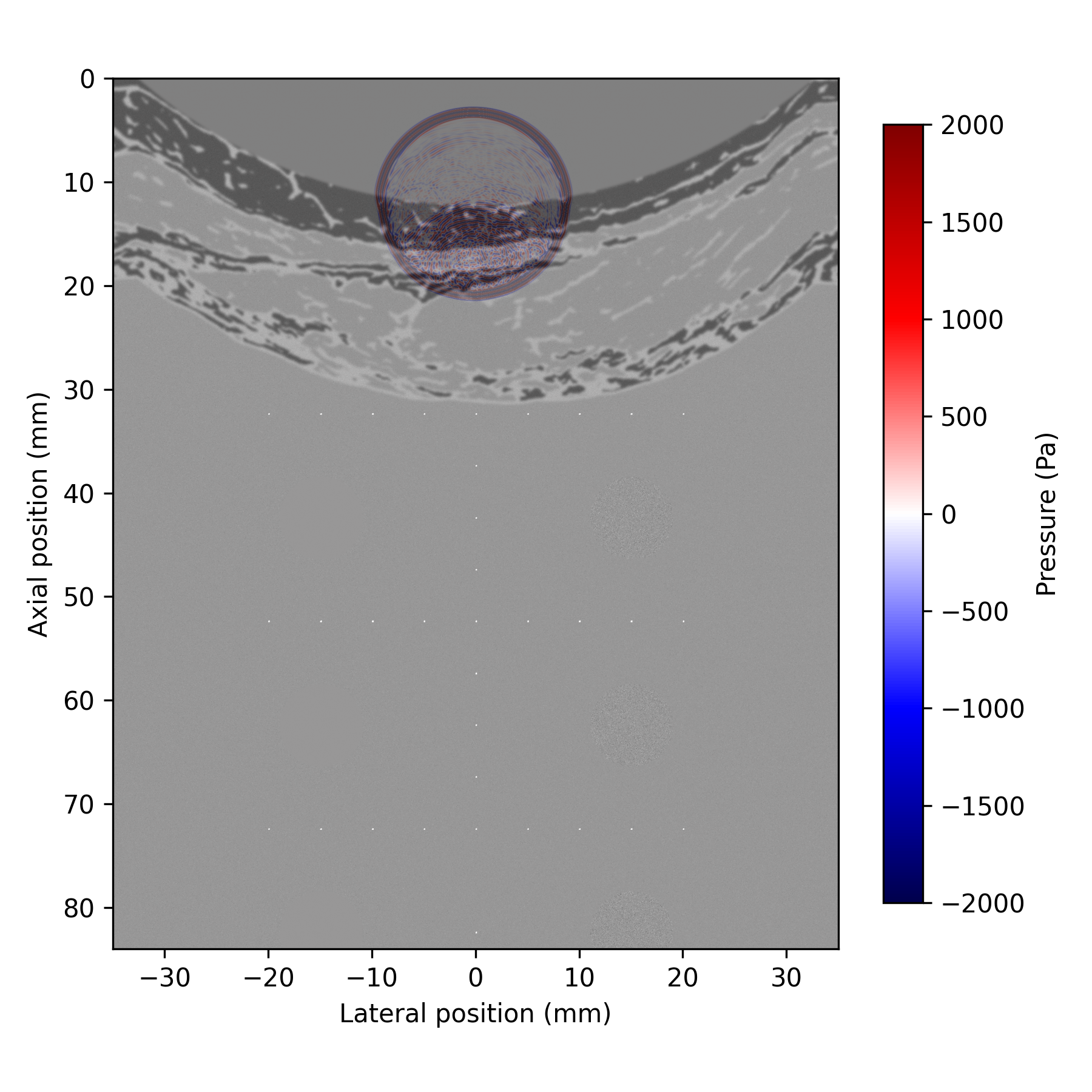}
            \subcaption{\(t = 6.31\) \textmu s}
        \end{minipage}
        \begin{minipage}[t]{0.28\textwidth}
            \centering
            \includegraphics[width=1.0\linewidth]{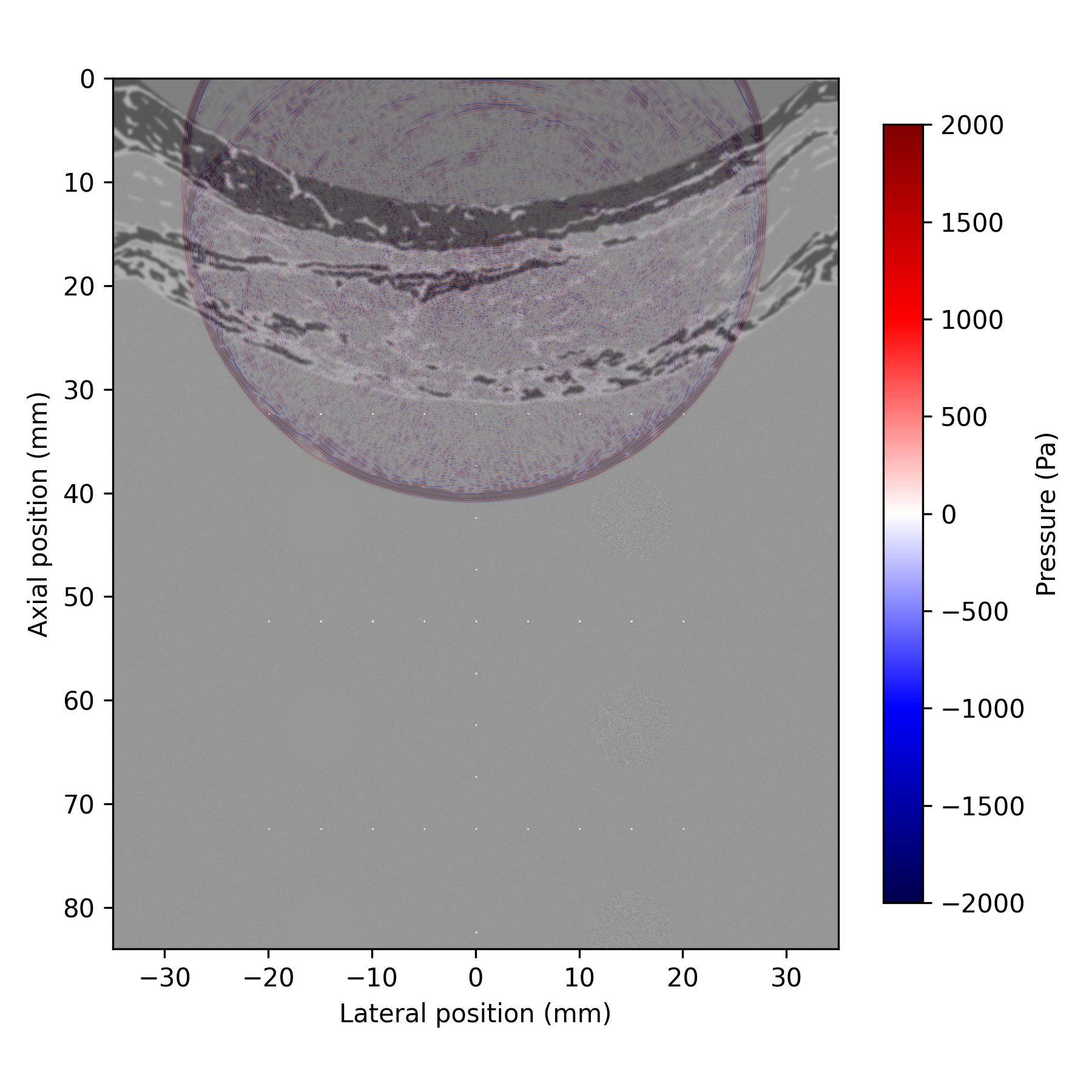}
            \subcaption{\(t = 18.9\) \textmu s}
        \end{minipage}
        \begin{minipage}[t]{0.28\textwidth}
            \centering
            \includegraphics[width=1.0\linewidth]{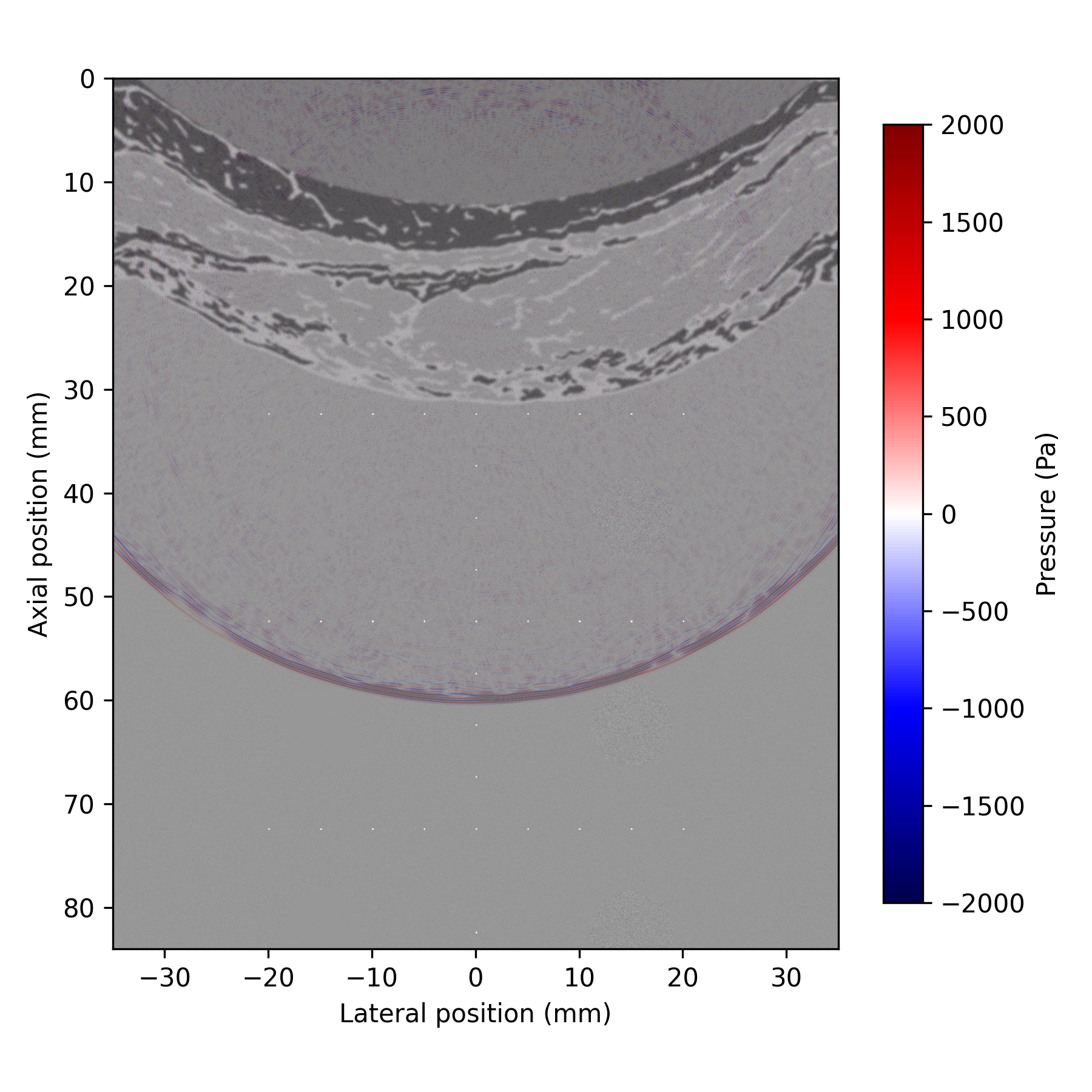}
            \subcaption{\(t = 31.5\) \textmu s}
        \end{minipage}
        \caption{Simulated wave field propagating through the abdominal wall phantom at different time points: (a) t = 6.31 \textmu s, (b) t = 18.9 \textmu s, and (c) t = 31.5 \textmu s. The wavefront distortion and amplitude attenuation caused by the heterogeneous abdominal wall tissues are clearly observed in the wave field.}
        \label{fig:simulated_abdominal_wavefield}
    \end{figure}
    \begin{figure}[t]
        \centering
        \begin{minipage}[t]{0.4\textwidth}
            \centering
            \includegraphics[width=1.0\linewidth]{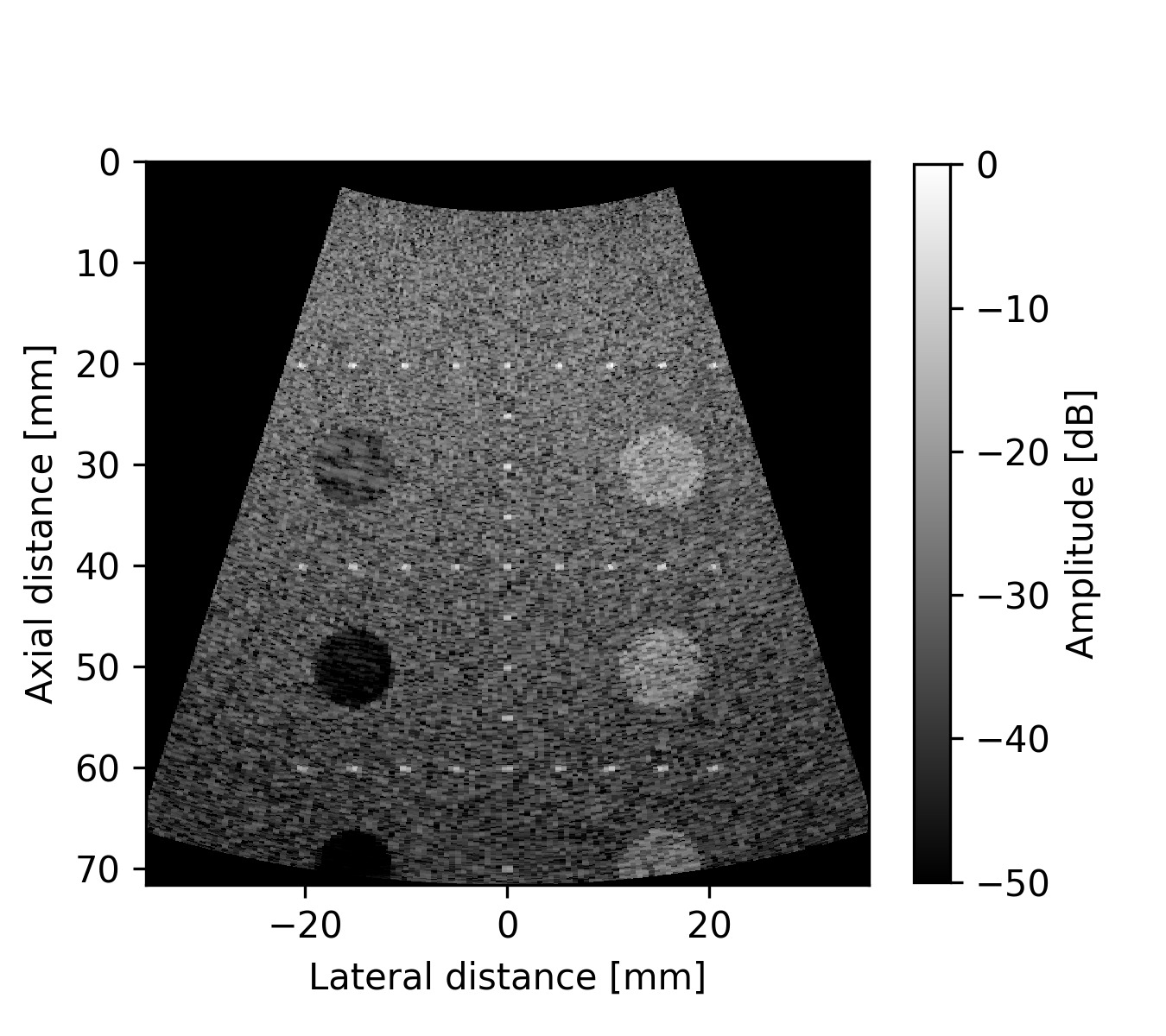}
            \subcaption{Homogeneous medium B-mode}
            \label{fig:simulated_bmode_c52v_homogeneous}

        \end{minipage}
        \begin{minipage}[t]{0.4\textwidth}
            \centering
            \includegraphics[width=1.0\linewidth]{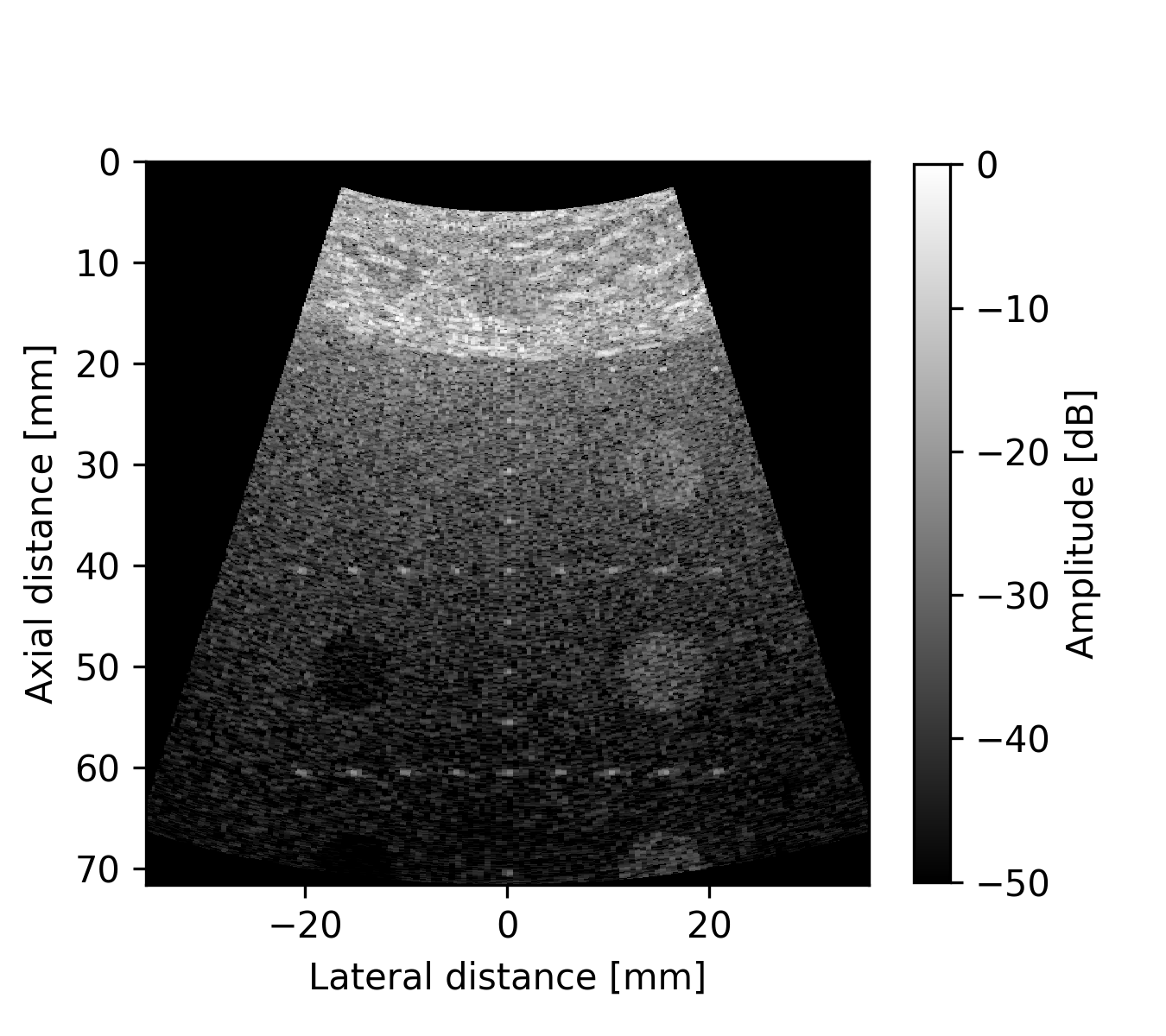}
            \subcaption{Abdominal phantom B-mode}
            \label{fig:simulated_bmode_c52v_heterogeneous}
        \end{minipage}
        \caption{simulated B-mode images with C5-2V transducer in (a) homogeneous medium and (b) abdominal wall phantom with heterogeneous attenuation properties. The image degradation due to the reverberation and aberration artifacts caused by the heterogeneous abdominal wall tissues is clearly observed in abdominal wall target B-mode image (b), while the simulated B-mode image in homogeneous medium (a) shows a clear image of the targets without artifacts.}
    \end{figure}
    \begin{figure}[t]
        \centering
        \begin{minipage}[t]{0.33\textwidth}
            \centering
            \includegraphics[width=1.0\linewidth]{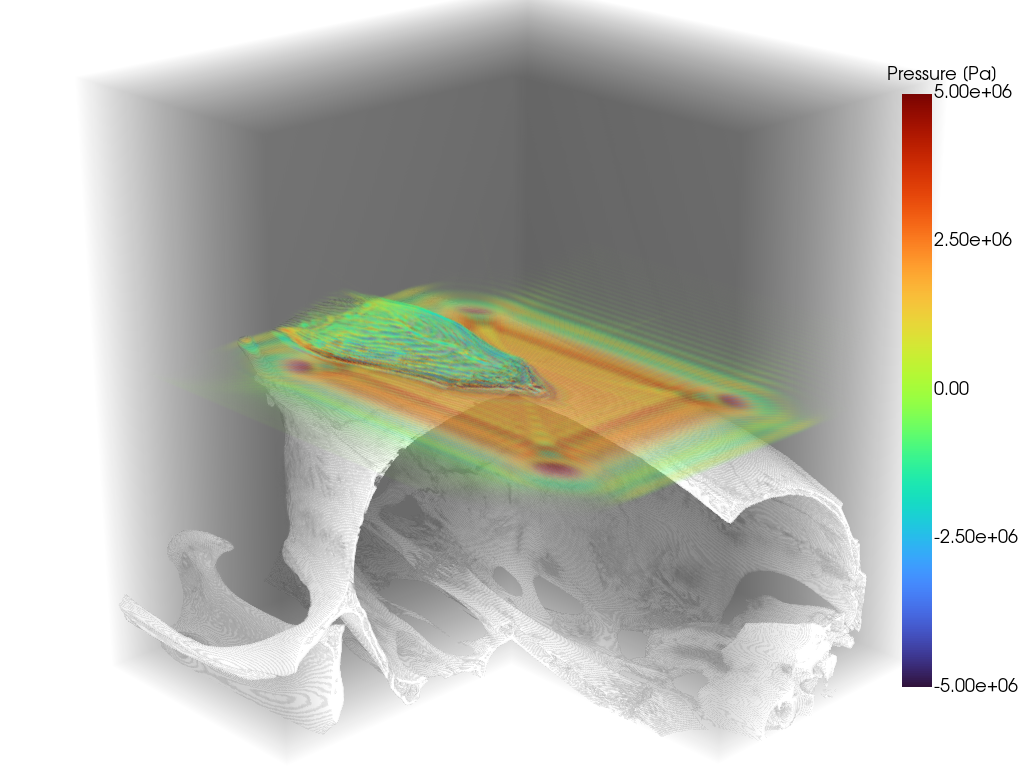}
            \subcaption{\(t = 1.07\) \textmu s}
        \end{minipage}
        \begin{minipage}[t]{0.33\textwidth}
            \centering
            \includegraphics[width=1.0\linewidth]{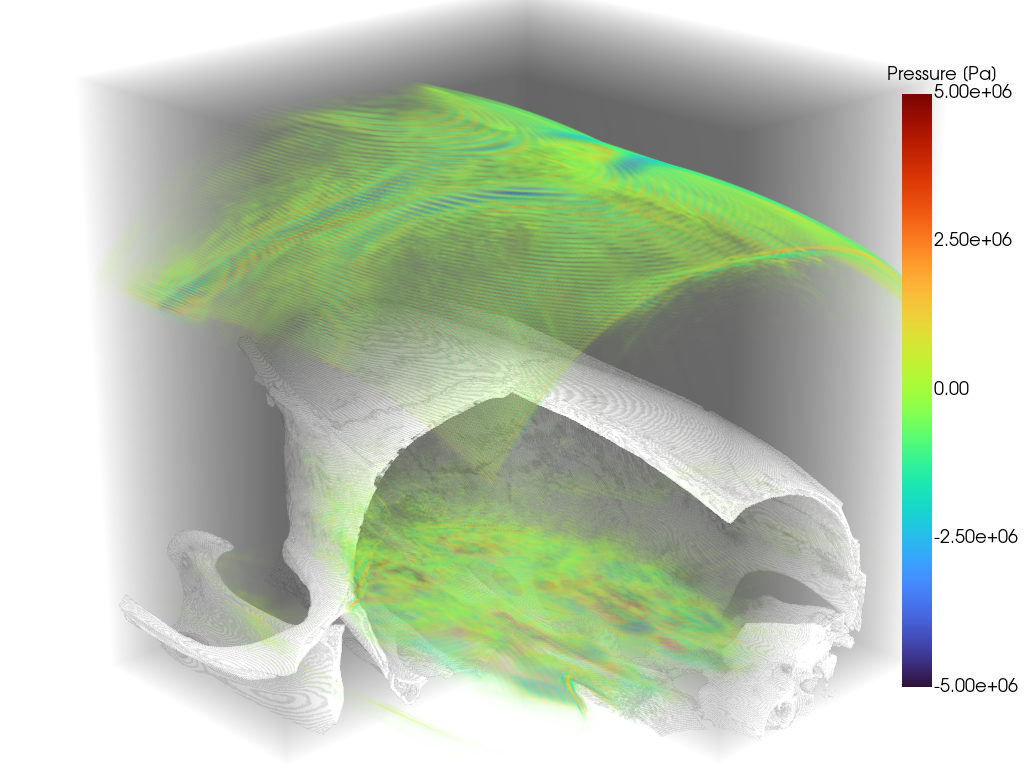}
            \subcaption{\(t = 1.86\) \textmu s}
        \end{minipage}
        \caption{Simulated wave field propagating through the rat skull phantom at different time points: (a) t = 1.07 \textmu s and (b) t = 1.86 \textmu s. The wavefront distortion and amplitude attenuation caused by the heterogeneous skull bone are clearly observed in the wave field. Part of the skull bone is not shown in the images to visualize the wave field inside the skull.}
        \label{fig:simulated_transcranial_wavefield}
    \end{figure}

    \section{2D Application: Abdominal wall imaging} \label{appendix:abdominal_wall_imaging}

    To demonstrate the practical application of the numerical implementation, we simulated B-mode ultrasound imaging of an abdominal wall phantom with heterogeneous attenuation properties.
    The abdominal wall phantom was modeled with layers representing skin, fat, muscle, and connective tissues, each assigned appropriate acoustic properties, including density, sound speed, and power-law attenuation characteristics.
    The phantom was constructed using a segmented Visible Human Project dataset \citep{Spitzer1996-ly, Ackerman1998-xp}.
    The segmentation process of the abdominal wall tissues was performed similarly to \citet{Zhuang2025-ym}.
    The acoustic properties for each tissue type were assigned based on literature values \citep{Lin1987-nn, Chen1987-bz, Sehgal1986-aj, Errabolu1987-du, Korta_Martiartu2021-lq, Agemura1990-oy, Edwards1988-pp, Youssef2018-up}.
    The power-law exponents were set to $1.0$ for all tissue types, while the attenuation coefficients $\alpha_0$ were set to $2.1 \ \text{dB/cm/MHz}$ for skin, $0.4 \ \text{dB/cm/MHz}$ for fat, $0.15 \ \text{dB/cm/MHz}$ for muscle, and $0.5 \ \text{dB/cm/MHz}$ for connective tissue.
    A convex transducer array (C5-2V, Verasonics Inc., Kirkland, WA, USA) was emulated in the simulation.
    We used the same settings as \citet{Zhuang2025-ym}.
    The center frequency was set to 3.7 MHz and a sampling frequency was set to 14.436 MHz.
    The transducer had 128 elements with a pitch of 0.508 mm and radius of 49.57 mm.
    A full synthetic aperture (FSA) sequence was performed by sequentially transmitting a pulse from each element and receiving the echoes on all elements. Figure \ref{fig:simulated_abdominal_wavefield} shows the simulated wave field propagating through the abdominal wall phantom at different time points.
    The wavefront distortion and amplitude attenuation caused by the heterogeneous abdominal wall tissues are clearly observed in the wave field.
    The simulation was performed using a spatial resolution of 12 points per wavelength (PPW) and a CFL number of 0.2.

    Figure \ref{fig:simulated_bmode_c52v_homogeneous} and \ref{fig:simulated_bmode_c52v_heterogeneous} show the simulated B-mode images for the homogeneous medium and the abdominal wall phantom, respectively.
    The images are compressed using logarithmic compression with a dynamic range of 50 dB after normalizing the envelope-detected signal with the maximum value.
    The image degradation due to the reverberation and aberration artifacts caused by the heterogeneous abdominal wall tissues is clearly observed in abdominal wall target B-mode image (Fig. \ref{fig:simulated_bmode_c52v_heterogeneous}), while the simulated B-mode image in homogeneous medium (Fig. \ref{fig:simulated_bmode_c52v_homogeneous}) shows a clear image of the targets without artifacts.

    \section{3D Application: Rat skull imaging} \label{appendix:rat_skull_imaging}

    To demonstrate the practical application of the numerical implementation in 3D, we simulated transcranial ultrasound imaging of a rat brain through the skull bone with heterogeneous attenuation properties.
    The rat skull phantom was modeled using micro-CT images of a rat skull \citep{Aubry2003-ep}.
    The segmentation process of the skull bone and brain tissues was performed similarly to \citet{DeRuiter2025-xh}. The bone maps were isolated in the CT scan and then scaled linearly to represent the speed of sound and density of the skull bone.
    The acoustic properties for each tissue type were assigned based on literature values \citep{Duck1990-gd}.
    The maximum sound speed of 2900 m/s, density of 2200 kg/m$^3$, and attenuation of 15 dB/cm/MHz were assigned to the skull bone, while the background medium was set to water properties with sound speed of 1540 m/s, density of 1000 kg/m$^3$, and attenuation of 0.0 dB/cm/MHz.
    A Vermon 32$\times$32 matrix array transducer (Vermon S.A., Tours, France) was emulated in the simulation.
    The center frequency was set to 7.81 MHz and a sampling frequency was set to 66.9 MHz.
    A single plane wave transmission was performed to demonstrate the wave propagation through the skull bone.
    The simulation was performed using a spatial resolution of 12 points per wavelength (PPW) and a CFL number of 0.2.
    Figure \ref{fig:simulated_transcranial_wavefield} shows the simulated wave field propagating through the rat skull phantom at different time points.
    The wavefront distortion and amplitude attenuation caused by the heterogeneous skull bone are clearly observed in the wave field.
    The computations were performed with CPU version Fullwave~2. The simulation took 69 hours on a server with 10 Intel Xeon Silver 4210R CPU cores.
    The visualization was performed using PyVista \citep{Sullivan2019-sb}.
\end{appendices}
     }
\fi

\bibliographystyle{plainnat}
\bibliography{references}

\ifarxiv
\begin{appendices}
    \section{Derivation of dispersion relation for the multiple relaxation model} \label{appendix:dispersion_relation}
    The derivation of the dispersion relation Eq.~(\ref{eq:dispersion_relations}) is provided below.
    Assuming that the governing equation is 1-dimensional plane wave, we obtain the solution $p(x, t) = p_0 e^{i (\omega t - kx)}$ and $v(x, t) = v_0 e^{i (\omega t - kx)}$. Taking the Fourier transform of Eq.~(\ref{eq:convolution_kernel_x}) yields,
    \begin{align}
        F\left[\zeta^\nu_{x_1}(t)\right] = -\cfrac{d^\nu_{x1}}{\kappa^2_{x1}} \left( \cfrac{1}{d_{x1}^\nu / \kappa_{x1} + \alpha_{x1}^{\nu} + i \omega} \right) = \tilde{\zeta}(\omega)
    \end{align}
    Considering convolution relationship in Fourier space and plane wave solution, the convolution term inside the summation in Eq. (\ref{eq:relaxation1}) for pressure will be
    \begin{align}
        F\left[\zeta(t) \ast p(x, t)\right] & = \tilde{\zeta}(\omega) \cdot F\left[p(x, t)\right]                                                                  \\
                                                  & = \tilde{\zeta}(\omega) \cdot \left[2 \pi p_0 e^{ikx} \delta(\omega - \omega_0)\right] \label{eq:fourier_plane_wave}
    \end{align}
    Taking an inverse Fourier transform of Eq.~(\ref{eq:fourier_plane_wave}) gives
    \begin{align}
        F^{-1}\left[\tilde{\zeta}(\omega) \cdot \left[2 \pi p_0 e^{ikx} \delta(\omega - \omega_0)\right]\right] & = \cfrac{1}{2 \pi} \int^\infty_{-\infty} \tilde{\zeta}(\omega) \cdot 2 \pi e ^ {ikx} \cdot e^{-i \omega t} \delta(\omega - \omega_0) d \omega             \\
                                                                                                                & = p_0 e^{ikx} \tilde{\zeta} (\omega_0) \int^{\infty}_{-\infty} \delta(\omega - \omega_0) e^{-i \omega t} d\omega                                          \\
                                                                                                                & =                                                                                 \tilde{\zeta}(\omega_0) \cdot p_0 e^{i (kx - \omega_0 t)}               \\
                                                                                                                & =                                                                                 \tilde{\zeta}(\omega_0) \cdot p(x, t) \label{eq:convolution_plane_wave}
    \end{align}
    Therefore, using Eq. (\ref{eq:gamma1}), Eq. (\ref{eq:convolution_plane_wave}), and Eq. (\ref{eq:relaxation1}),
    \begin{align}
        \cfrac{\partial p}{\partial \tilde{x}_1} & = \cfrac{1}{\kappa_{x_1}} \cdot \cfrac{\partial p}{\partial x} + \sum_{\nu=1}^{N} \zeta_{x_1}^\nu \ast \cfrac{\partial p}{\partial x} \\
                                                 & =  \cfrac{1}{\kappa_{x_1}} \cdot \cfrac{\partial p}{\partial x} + \sum^{N}_{\nu=1} \tilde{\zeta}(\omega) (p ik)                       \\
                                                 & = \cfrac{1}{\kappa_{x_1}} \cdot ikp - \gamma_1 \cdot ikp                                                                              \\
                                                 & = ikp \left[\cfrac{1}{\kappa_{x_1}} - \gamma_1\right] \label{eq:relaxation_derivative}
    \end{align}
    substituting Eq.~(\ref{eq:relaxation_derivative}) to (\ref{eq:wave1}), we obtain
    \begin{align}
        ik \left(\cfrac{1}{\kappa_{x_1}} - \gamma_1\right) p - i\omega \rho v = 0 \\
        \therefore \ v = \cfrac{k}{\omega \rho} \left(\cfrac{1}{\kappa_{x_1}} - \gamma_1\right) p \label{eq:wave1_velocity}
    \end{align}
    substituting Eq.~(\ref{eq:relaxation_derivative}) to (\ref{eq:wave2}), where $K = 1/(\rho c^2)$ is the medium compressibility, we obtain
    \begin{align}
        \nabla_2 \cdot \matr{v} & = \left(\cfrac{1}{\kappa_{x_2}} - \gamma_2\right) \cdot ik\matr{v}                                 \\
        \therefore v            & = \cfrac{\omega K p}{k \left(\cfrac{1}{\kappa_{x_2}} - \gamma_2 \right)} \label{eq:wave2_velocity}
    \end{align}
    Equating Eq. (\ref{eq:wave1_velocity}) and Eq. (\ref{eq:wave2_velocity}), we obtain the dispersion relation Eq. (\ref{eq:dispersion_relations}).

    \begin{align}
        k^2 \left(\cfrac{1}{\kappa_{x_1}} - \gamma_1\right) \left(\cfrac{1}{\kappa_{x_2}} - \gamma_2\right)  = \omega ^ 2 \rho K
        = \cfrac{\omega ^ 2}{c ^ 2} \\
        \therefore k = \cfrac{\omega}{c} \left( \cfrac{1}{\kappa_{x1} \kappa_{x2}} - \cfrac{\gamma_1}{\kappa_{x2}} - \cfrac{\gamma_2}{\kappa_{x1}} + \gamma_1 \gamma_2 \right) ^ {-\frac{1}{2}}
    \end{align}
    \begin{figure}[b]
        \centering
        \begin{minipage}[t]{0.28\textwidth}
            \centering
            \includegraphics[width=1.0\linewidth]{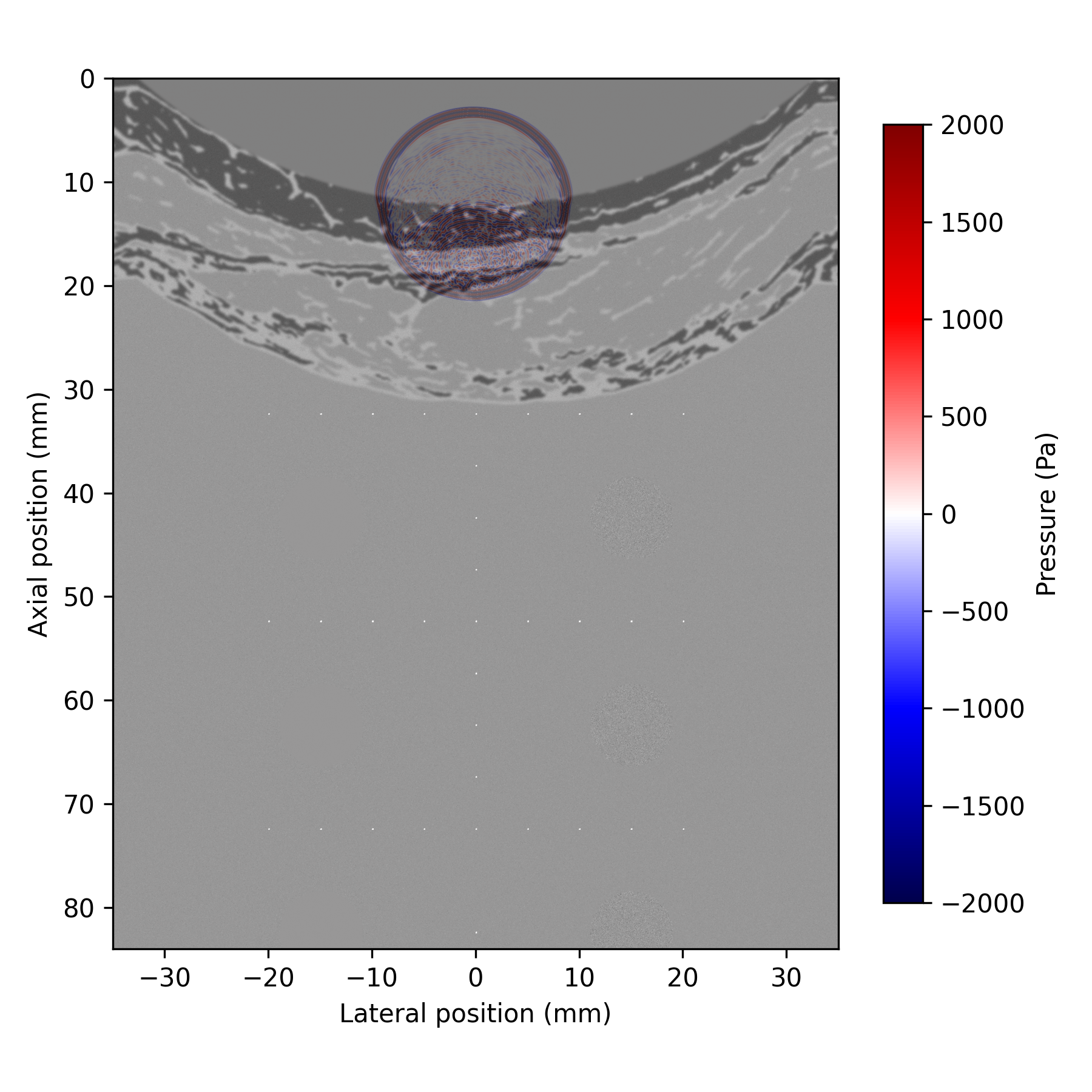}
            \subcaption{\(t = 6.31\) \textmu s}
        \end{minipage}
        \begin{minipage}[t]{0.28\textwidth}
            \centering
            \includegraphics[width=1.0\linewidth]{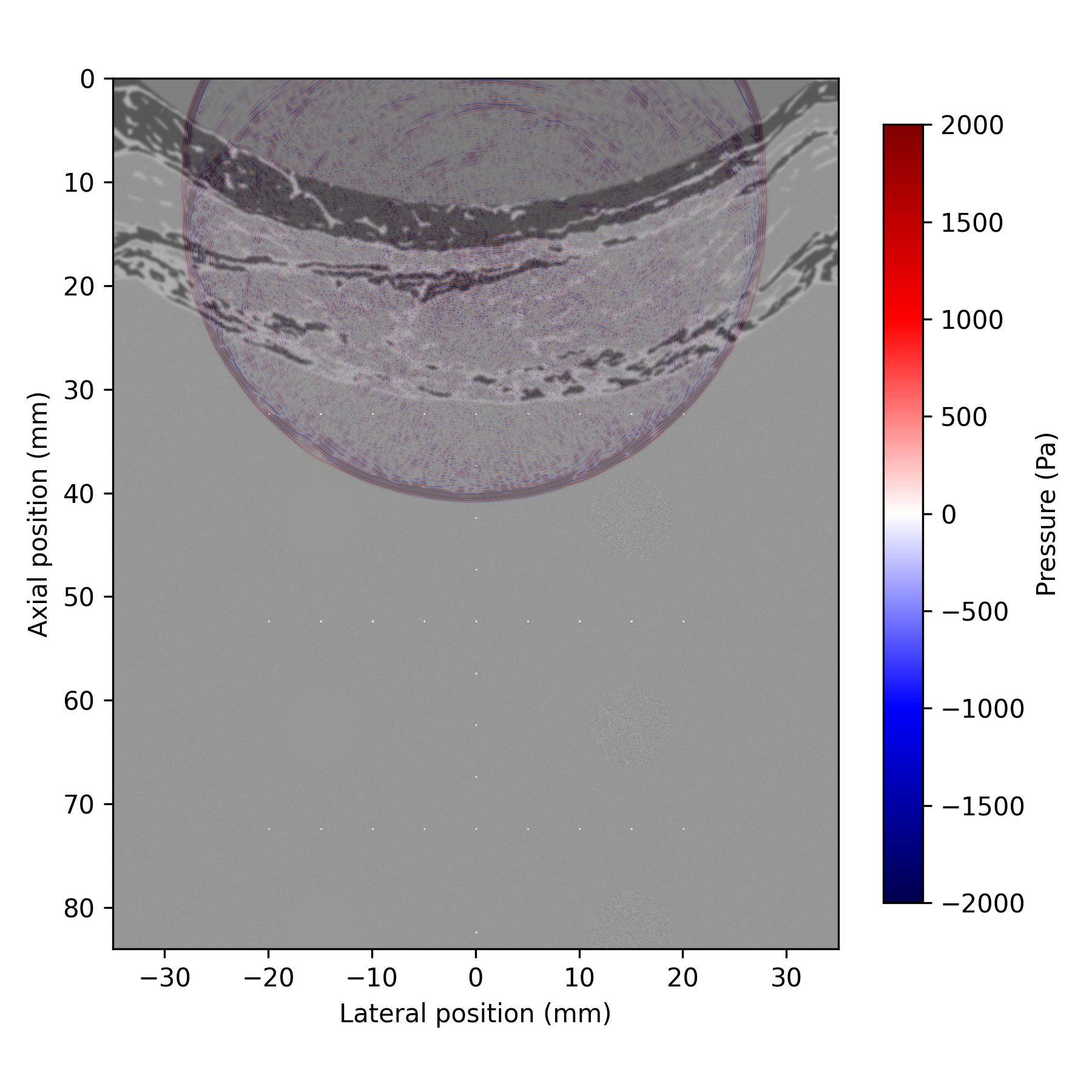}
            \subcaption{\(t = 18.9\) \textmu s}
        \end{minipage}
        \begin{minipage}[t]{0.28\textwidth}
            \centering
            \includegraphics[width=1.0\linewidth]{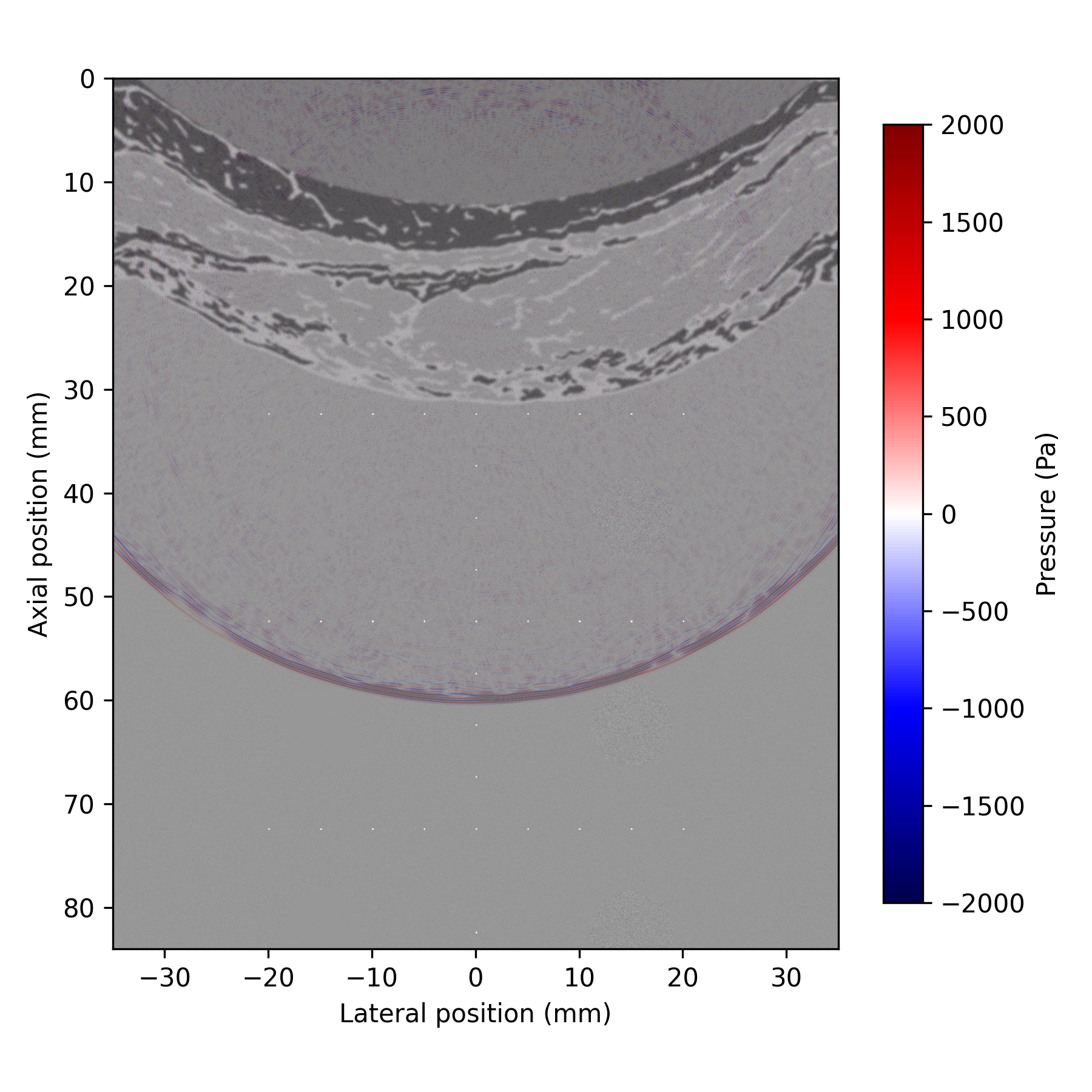}
            \subcaption{\(t = 31.5\) \textmu s}
        \end{minipage}
        \caption{Simulated wave field propagating through the abdominal wall phantom at different time points: (a) t = 6.31 \textmu s, (b) t = 18.9 \textmu s, and (c) t = 31.5 \textmu s. The wavefront distortion and amplitude attenuation caused by the heterogeneous abdominal wall tissues are clearly observed in the wave field.}
        \label{fig:simulated_abdominal_wavefield}
    \end{figure}
    \begin{figure}[t]
        \centering
        \begin{minipage}[t]{0.4\textwidth}
            \centering
            \includegraphics[width=1.0\linewidth]{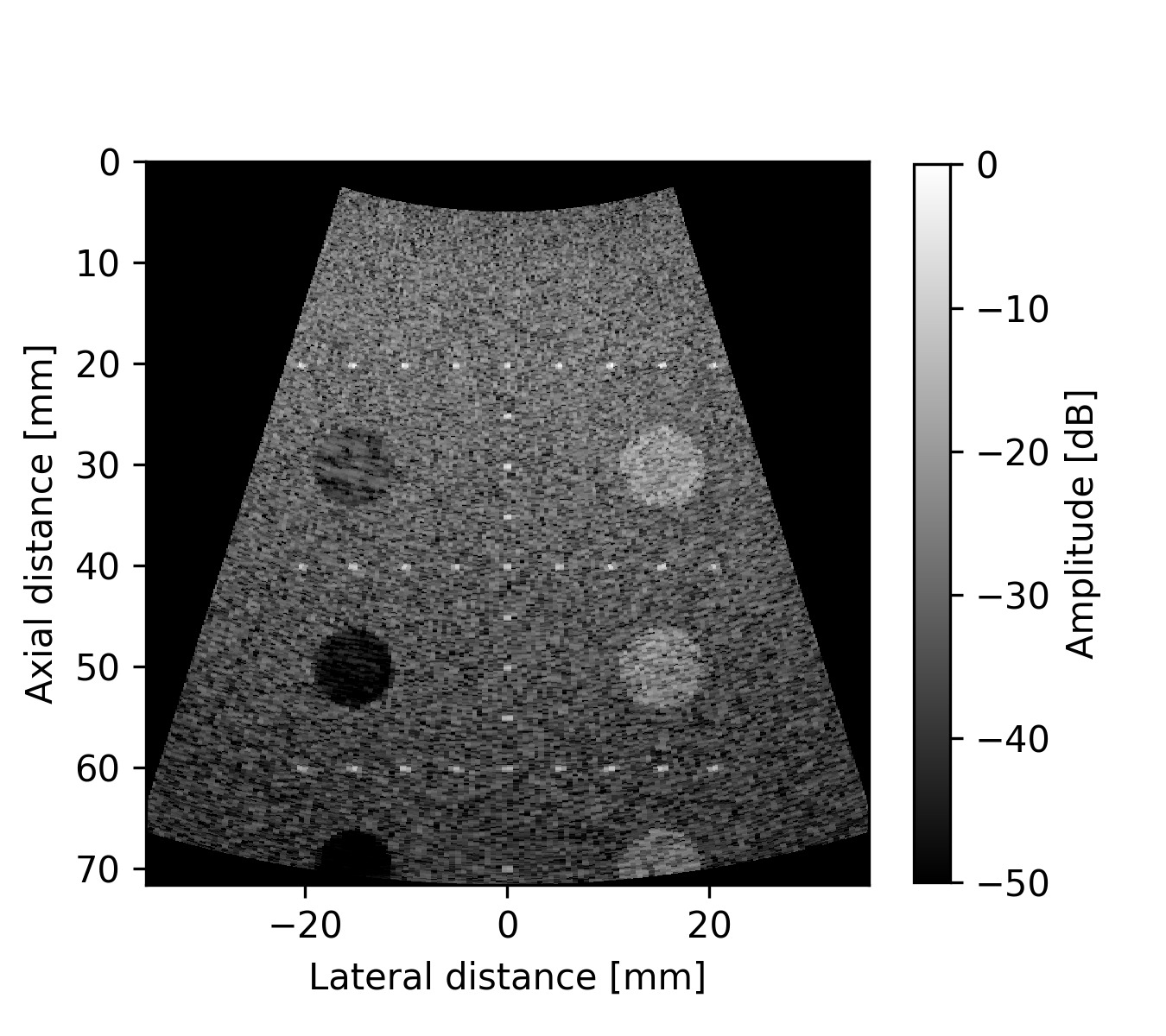}
            \subcaption{Homogeneous medium B-mode}
            \label{fig:simulated_bmode_c52v_homogeneous}

        \end{minipage}
        \begin{minipage}[t]{0.4\textwidth}
            \centering
            \includegraphics[width=1.0\linewidth]{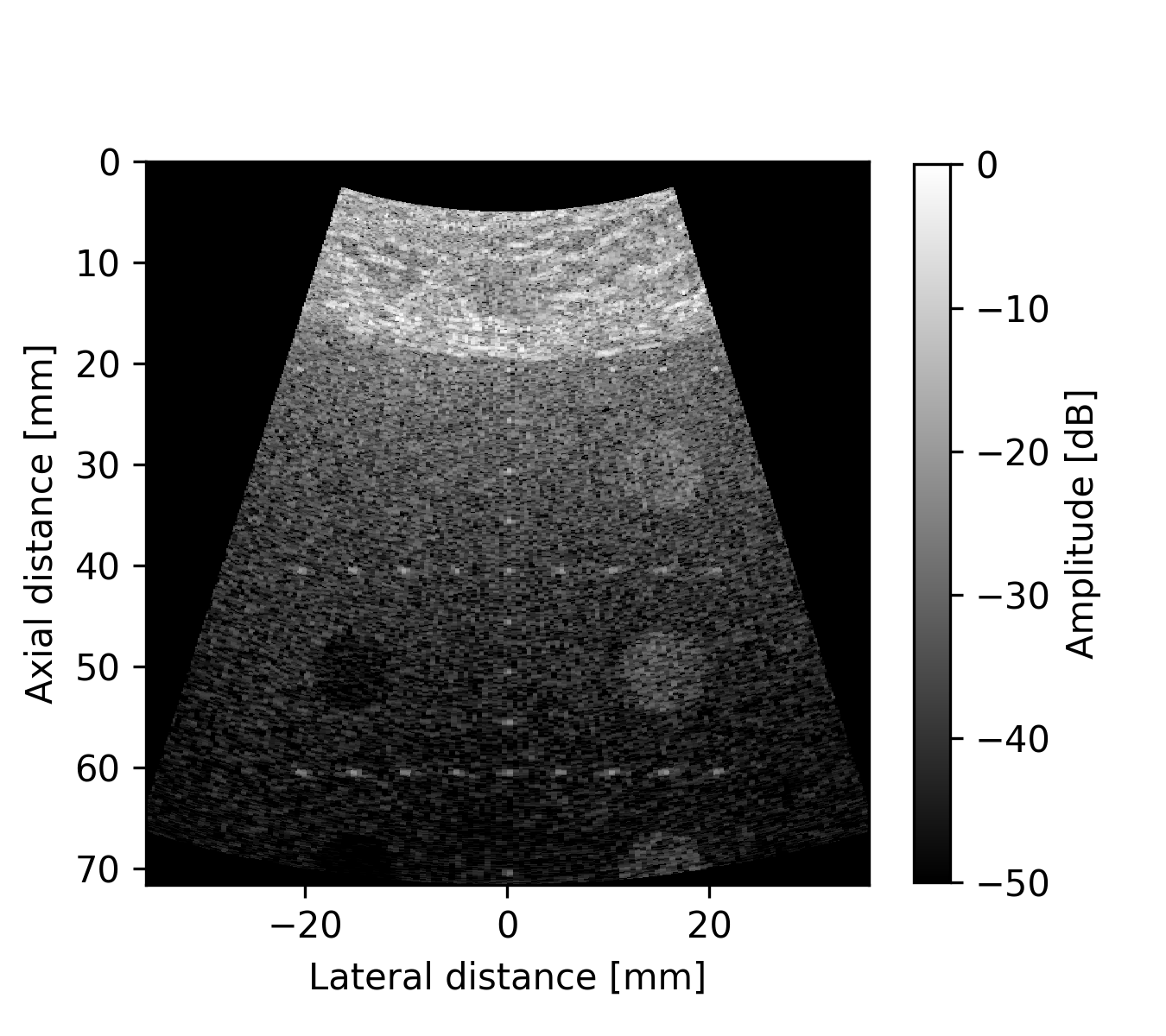}
            \subcaption{Abdominal phantom B-mode}
            \label{fig:simulated_bmode_c52v_heterogeneous}
        \end{minipage}
        \caption{simulated B-mode images with C5-2V transducer in (a) homogeneous medium and (b) abdominal wall phantom with heterogeneous attenuation properties. The image degradation due to the reverberation and aberration artifacts caused by the heterogeneous abdominal wall tissues is clearly observed in abdominal wall target B-mode image (b), while the simulated B-mode image in homogeneous medium (a) shows a clear image of the targets without artifacts.}
    \end{figure}
    \begin{figure}[t]
        \centering
        \begin{minipage}[t]{0.33\textwidth}
            \centering
            \includegraphics[width=1.0\linewidth]{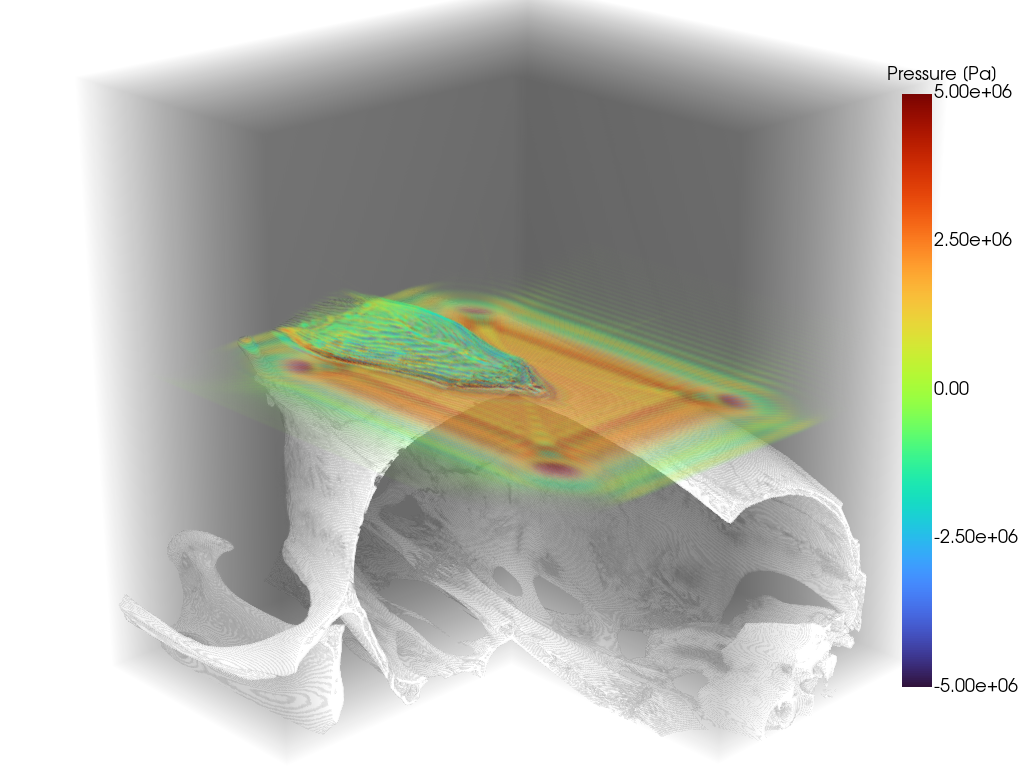}
            \subcaption{\(t = 1.07\) \textmu s}
        \end{minipage}
        \begin{minipage}[t]{0.33\textwidth}
            \centering
            \includegraphics[width=1.0\linewidth]{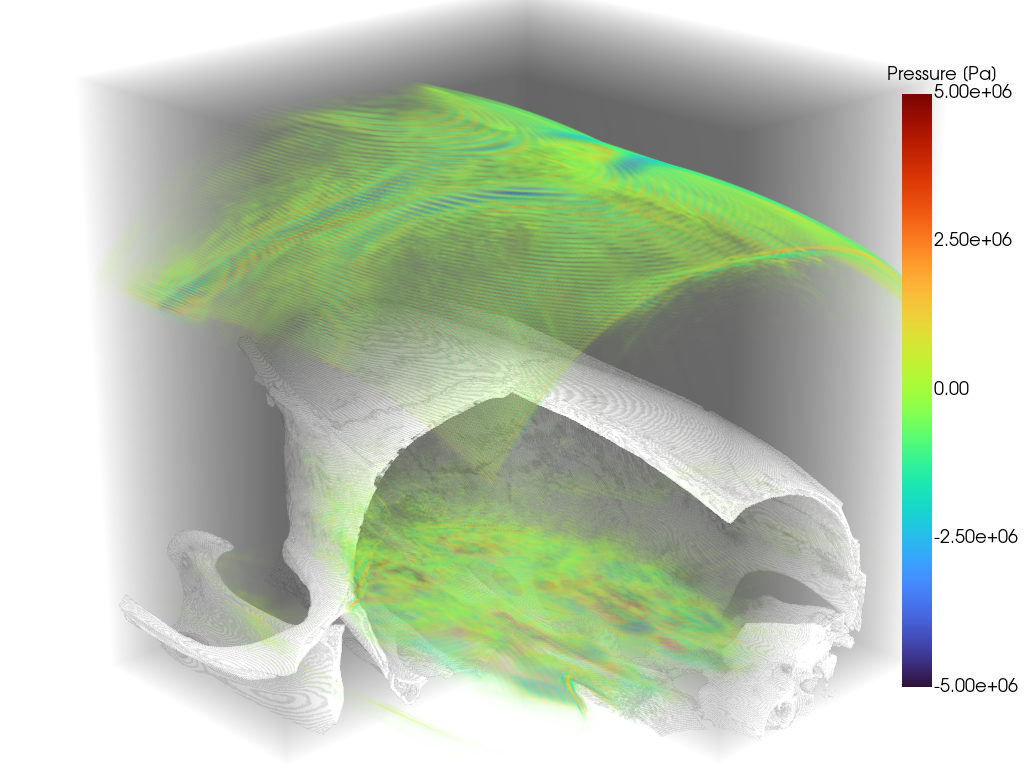}
            \subcaption{\(t = 1.86\) \textmu s}
        \end{minipage}
        \caption{Simulated wave field propagating through the rat skull phantom at different time points: (a) t = 1.07 \textmu s and (b) t = 1.86 \textmu s. The wavefront distortion and amplitude attenuation caused by the heterogeneous skull bone are clearly observed in the wave field. Part of the skull bone is not shown in the images to visualize the wave field inside the skull.}
        \label{fig:simulated_transcranial_wavefield}
    \end{figure}

    \section{2D Application: Abdominal wall imaging} \label{appendix:abdominal_wall_imaging}

    To demonstrate the practical application of the numerical implementation, we simulated B-mode ultrasound imaging of an abdominal wall phantom with heterogeneous attenuation properties.
    The abdominal wall phantom was modeled with layers representing skin, fat, muscle, and connective tissues, each assigned appropriate acoustic properties, including density, sound speed, and power-law attenuation characteristics.
    The phantom was constructed using a segmented Visible Human Project dataset \citep{Spitzer1996-ly, Ackerman1998-xp}.
    The segmentation process of the abdominal wall tissues was performed similarly to \citet{Zhuang2025-ym}.
    The acoustic properties for each tissue type were assigned based on literature values \citep{Lin1987-nn, Chen1987-bz, Sehgal1986-aj, Errabolu1987-du, Korta_Martiartu2021-lq, Agemura1990-oy, Edwards1988-pp, Youssef2018-up}.
    The power-law exponents were set to $1.0$ for all tissue types, while the attenuation coefficients $\alpha_0$ were set to $2.1 \ \text{dB/cm/MHz}$ for skin, $0.4 \ \text{dB/cm/MHz}$ for fat, $0.15 \ \text{dB/cm/MHz}$ for muscle, and $0.5 \ \text{dB/cm/MHz}$ for connective tissue.
    A convex transducer array (C5-2V, Verasonics Inc., Kirkland, WA, USA) was emulated in the simulation.
    We used the same settings as \citet{Zhuang2025-ym}.
    The center frequency was set to 3.7 MHz and a sampling frequency was set to 14.436 MHz.
    The transducer had 128 elements with a pitch of 0.508 mm and radius of 49.57 mm.
    A full synthetic aperture (FSA) sequence was performed by sequentially transmitting a pulse from each element and receiving the echoes on all elements. Figure \ref{fig:simulated_abdominal_wavefield} shows the simulated wave field propagating through the abdominal wall phantom at different time points.
    The wavefront distortion and amplitude attenuation caused by the heterogeneous abdominal wall tissues are clearly observed in the wave field.
    The simulation was performed using a spatial resolution of 12 points per wavelength (PPW) and a CFL number of 0.2.

    Figure \ref{fig:simulated_bmode_c52v_homogeneous} and \ref{fig:simulated_bmode_c52v_heterogeneous} show the simulated B-mode images for the homogeneous medium and the abdominal wall phantom, respectively.
    The images are compressed using logarithmic compression with a dynamic range of 50 dB after normalizing the envelope-detected signal with the maximum value.
    The image degradation due to the reverberation and aberration artifacts caused by the heterogeneous abdominal wall tissues is clearly observed in abdominal wall target B-mode image (Fig. \ref{fig:simulated_bmode_c52v_heterogeneous}), while the simulated B-mode image in homogeneous medium (Fig. \ref{fig:simulated_bmode_c52v_homogeneous}) shows a clear image of the targets without artifacts.

    \section{3D Application: Rat skull imaging} \label{appendix:rat_skull_imaging}

    To demonstrate the practical application of the numerical implementation in 3D, we simulated transcranial ultrasound imaging of a rat brain through the skull bone with heterogeneous attenuation properties.
    The rat skull phantom was modeled using micro-CT images of a rat skull \citep{Aubry2003-ep}.
    The segmentation process of the skull bone and brain tissues was performed similarly to \citet{DeRuiter2025-xh}. The bone maps were isolated in the CT scan and then scaled linearly to represent the speed of sound and density of the skull bone.
    The acoustic properties for each tissue type were assigned based on literature values \citep{Duck1990-gd}.
    The maximum sound speed of 2900 m/s, density of 2200 kg/m$^3$, and attenuation of 15 dB/cm/MHz were assigned to the skull bone, while the background medium was set to water properties with sound speed of 1540 m/s, density of 1000 kg/m$^3$, and attenuation of 0.0 dB/cm/MHz.
    A Vermon 32$\times$32 matrix array transducer (Vermon S.A., Tours, France) was emulated in the simulation.
    The center frequency was set to 7.81 MHz and a sampling frequency was set to 66.9 MHz.
    A single plane wave transmission was performed to demonstrate the wave propagation through the skull bone.
    The simulation was performed using a spatial resolution of 12 points per wavelength (PPW) and a CFL number of 0.2.
    Figure \ref{fig:simulated_transcranial_wavefield} shows the simulated wave field propagating through the rat skull phantom at different time points.
    The wavefront distortion and amplitude attenuation caused by the heterogeneous skull bone are clearly observed in the wave field.
    The computations were performed with CPU version Fullwave~2. The simulation took 69 hours on a server with 10 Intel Xeon Silver 4210R CPU cores.
    The visualization was performed using PyVista \citep{Sullivan2019-sb}.
\end{appendices}
 \fi

\end{document}